\documentclass[prb,twocolumn]{revtex4-1}%
\usepackage[dvips,final]{graphicx}
\usepackage{epsfig}
\usepackage{amsmath,amsfonts,amssymb,bm}
\usepackage{float}
\usepackage[caption = false]{subfig}
\usepackage{multirow}
\setcitestyle{numbers,square}

\begin{document}
\title{Theory of nonlinear microwave absorption by interacting two-level systems}
\author{Alexander L. Burin, Andrii O. Maksymov} 
\affiliation{Tulane University, New
Orleans, LA 70118, USA}

\date{\today}
\begin{abstract}
The microwave absorption and noise caused by quantum two-level systems (TLS) dramatically suppress the coherence in Josephson junction qubits that are promising candidates for  quantum information applications. It is a challenge to understand microwave absorption by TLSs because of the spectral diffusion resulting from fluctuations in their resonant frequencies induced by  their  long-range interactions.
Here we treat the spectral diffusion  explicitly  using the generalized master equation formalism. The proposed theory predicts that the linear absorption regime holds while a TLS Rabi frequency is smaller than  their phase decoherence rate. At higher external fields, a novel non-linear absorption regime is found with the loss tangent inversely proportional to the intensity of the field. The theory can be generalized to acoustic absorption and lower dimensions realized in superconducting qubits. 
\end{abstract}

\maketitle

\section{Introduction}

\begin{figure}[h!]
\centering
\includegraphics[width=\columnwidth]{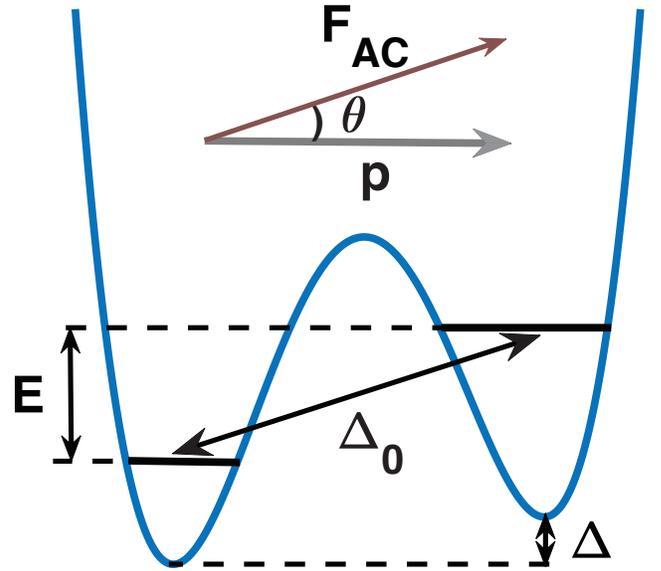}
\caption{A tunneling two-level system having an energy splitting $E$ and a dipole moment $\mathbf{p}$ interacting with an external field $\mathbf{F}_{AC}\cos(\omega t)$. The energy $E=\sqrt{\Delta^2+\Delta_{0}^2}$  is defined by an asymmetry $\Delta$ and a tunneling amplitude $\Delta_0$ \cite{AHV,Ph}.}
\label{fig:Fig1TLS}
\end{figure}

Quantum two-level systems (TLSs),  commonly represented by atoms or groups of atoms tunneling between two states (see Fig. \ref{fig:Fig1TLS}, Refs. \cite{AHV,Ph}), are ubiquitous in amorphous solids. 
TLSs restrict performance of modern nanodevices including  superconducting qubits \cite{Martinis05} and quantum dots \cite{KuhlmannNoiseInQDots2013} for quantum computing, kinetic inductance photon detectors for astronomy \cite{Gao08} and nanomechanical resonators \cite{Hauer17}. TLSs are commonly found in Josephson junction barriers, wiring crossovers \cite{PaikOsborn10,UstinovScience12,LisenfeldReview17} and even on the surfaces of the resonators with coplanar superconducting electrodes on crystalline substrates \cite{Gao08,BurnettFaoro14}.  TLSs reduce the coherence in qubits absorbing microwaves in the frequency domain of their oscillations \cite{Martinis05,LisenfeldReview17} and 
producing a noise in qubit resonant energies  \cite{Yu04,Yu09,BurnettFaoro14,FaoroIoffe15,ab15TLSnoise,GalperinYuLZFlickerNoiseRev14}.

Both the microwave absorption and the noise induced by TLSs are dramatically sensitive to their interactions 
 \cite{Hunklinger77NonLin,BurnettFaoro14,GalperinYuLZFlickerNoiseRev14,FaoroIoffe15,ab15TLSnoise,
FaoroIoffe12,Weiss17}. While the $1/f$ noise is reasonably interpreted within the interacting TLS model \cite{FaoroIoffe15,ab15TLSnoise}, the problem of non-linear microwave absorption  is not resolved yet, in spite of the numerous efforts \cite{Hunklinger77NonLin,Enss02Rev,FaoroIoffe12}. Here we propose a solution to this long-standing problem integrating the earlier developed rate equation model for TLS density matrix time evolution \cite{HunklingerReview,Yu94} briefly introduced in Sec. \ref{sec:RateEq} with the master equation formalism to account for the spectral diffusion \cite{Stratonovichbook,LaikhtmanSD86,GalperinYuLZTelNoise} introduced in Sec. \ref{sec:masteqDer}.  
The rigorous solution of the master equation is obtained in Sec. \ref{sec:Solution} for  the low temperature case where the thermal energy is smaller than the  microwave energy quantum 
\begin{eqnarray}
k_{B}T < \hbar\omega.
\label{eq:upperlim}
\end{eqnarray} 
Based on this solution the analytical expression for the loss tangent is derived in Sec. \ref{sec:lt}. The opposite high temperature regime is discussed qualitatively in Sec. \ref{sec:ltoutscope}. 
The obtained solution results in the novel  behaviors of non-linear absorption in a qualitative agreement with earlier expectations \cite{LaikhtmanSD86,Galperin88}. The predictions of theory are partially consistent with the experimental observations in amorphous solids  \cite{Hunklinger77NonLin} and deviate from the observed weakening of intensity dependence in Josephson junction resonators and qubits. The possible reasons for deviations are discussed in Sec. \ref{sec:exp}. The paper is resumed by the brief conclusion in Sec. \ref{sec:concl} including the summary of loss tangent non-linear behaviors at different temperatures given in Table \ref{tbl:summary}. 




\section{Rate equation formalism and TLS loss tangent}
\label{sec:RateEq}

The non-linear absorption of acoustic and electromagnetic waves by TLSs in amorphous solids was discovered experimentally almost half a century ago  \cite{Hunklinger72NonlinAbs,Golding73,Golding76,Hunklinger77NonLin}. The results have been 
interpreted using the rate equation formalism  applied to the TLS density matrix  \cite{Hunklinger77NonLin,Yu94} in the Bloch vector representation \cite{GalperinYuLZTelNoise} defined as $\sigma^{x}=\rho_{ge}+\rho_{eg}$, 
$\sigma^{y}=(\rho_{ge}-\rho_{eg})/i$, $\sigma^{z}=\rho_{gg}-\rho_{ee}$, where indices $g$ and $e$ stand for the TLS ground and excited states, respectively. 
The time evolution of the density matrix is determined by TLS frequency detuning from resonance $D=E/\hbar-\omega$, Rabi frequency $\Omega_{R}=(\Delta_{0}/E) pF_{AC}\cos(\theta)$ (see Fig. \ref{fig:Fig1TLS}), and relaxation and decoherence times $T_{1}$ and $T_{2}$. In the rotating frame approximation, relevant in the regime of interest, Eq. (\ref{eq:upperlim}), corresponding to the resonant absorption \cite{HunklingerReview,Yu94,ab95DipGap,ab98book}, one has 
\begin{eqnarray}
\frac{d\sigma^{x}}{dt}=D \sigma^{y}-\frac{\sigma^{x}}{T_{2}}, ~ \frac{d\sigma^{y}}{dt}=-D \sigma^{x}-\frac{\sigma^{y}}{T_{2}}+\Omega_{R}\sigma^{z}, 
\nonumber\\
\frac{d\sigma^{z}}{dt}=-\Omega_{R}\sigma^{y}-\frac{\sigma^{z}-\sigma^{z}_{eq}}{T_{1}}, ~ \sigma^{z}_{eq}=\tanh\left(\frac{E}{2k_{B}T}\right). 
\label{eq:BlochBas}
\end{eqnarray} 
The equilibrium population difference $\sigma^{z}_{eq}$ for resonant TLSs ($E\approx \hbar\omega$) can be set equal to unity in the case of interest $k_{B}T \ll \hbar\omega$. 
The rotating frames approximation is applicable at sufficiently small external field, $\Omega_{R}\ll \omega$, which is satisfied in the vast majority of measurements of TLS resonant absorption at microwave frequencies.


The TLS relaxation and decoherence times originated from the TLS interaction with phonons are defined as \cite{HunklingerReview,ab13LZTh,ab13echo,ab15TLSnoise}
\begin{eqnarray}
\frac{1}{T_{1}}= \alpha \frac{\Delta_{0}^2E}{k_{B}^3}  \coth\left(\frac{E}{2k_{B}T}\right) \approx \frac{x^2}{T_{10}}, ~ T_{2}=2T_{1}, 
\label{eq:TLSPars}
\end{eqnarray}
where $T_{10}$ stands for the minimum relaxation time and we introduced the dimensionless parameter $x=\Delta_{0}/E$ to describe TLS tunneling coupling. TLSs obey the universal distribution $P(E, x)=P_{0}/(x\sqrt{1-x^2})$ (see Fig. \ref{fig:Fig1TLS} and Refs.  \cite{AHV,Ph,PhillipsReview,HunklingerReview}). 
 The minimum decoherence time is given by $T_{20}=2T_{10}$ and the constant $\alpha \sim 3\cdot 10^{7}$ s$^{-1}$K$^{-3}$ \cite{ab13LZTh} is determined by the TLS-phonon interaction.

The microwave absorption is usually characterized by the TLS loss tangent  defined in terms of the integrated reactive response $\Omega_{R}\sigma^{y}(D)$. This response is  determined by the stationary solution of Eq. (\ref{eq:BlochBas}). 
For the sake of simplicity, we  assume all TLS dipole moments to have identical  absolute values, $p$, as argued in Refs. \cite{Martinis05,ab13LZTh}. The Rabi frequency can then be expressed in terms of the maximum Rabi frequency $\Omega_{R0}=pF_{AC}/\hbar$ as $\Omega_{R}=xy\Omega_{R0}$ where  $y=\cos(\theta)$ and it is uniformly distributed between $-1$ and $1$. 
 The loss tangent is defined as \cite{Yu94,ab13LZTh}
\begin{eqnarray}
\tan(\delta)=\frac{4\pi P_{0}\hbar^2}{2\epsilon'F_{AC}^2} \int_{-1}^{1}dy\int_{0}^{1} dx\frac{xy\Omega_{R0}f^{y}(0, x, y)}{x\sqrt{1-x^2}}, 
\nonumber\\
 f^{a}(q, x, y)=\int_{-\infty}^{\infty}\sigma^{a}(D, x, y)e^{iqD}dD, ~ a=x, y, z.
\label{eq:lossdefBas}
\end{eqnarray}
where $\epsilon'$ is a static dielectric constant of the material. 
The Fourier transforms $f^{a}(q, x, y)$ will be used in the future analysis. 

If the TLS Bloch vector obeys Eq. (\ref{eq:BlochBas}) then the Fourier transform of its $y$ component contributing to the loss tangent can be evaluated as (cf. Refs. \cite{Hunklinger77NonLin,Yu94})
\begin{eqnarray}
f^{y}(0) = \frac{\pi\Omega_{R}\tanh\left(\frac{\hbar\omega}{2k_{B}T}\right)}{\sqrt{1+\Omega_{R}^2T_{1}T_{2}}}.  
\label{eq:KummerSol3a0}
\end{eqnarray}
This steady state solution of Eq. (\ref{eq:BlochBas}) 
substituted into Eq. (\ref{eq:lossdefBas}) determines a  TLS loss tangent  in the approximate form  \cite{Hunklinger77NonLin,Golding83,Martinis05,Gao08,ab13LZTh}
\begin{eqnarray}
\tan(\delta)\approx\frac{\tan(\delta_{0})}{\sqrt{1+\frac{32}{9\pi^2}\Omega_{R0}^2 T_{20}^2}},
\nonumber\\
\tan(\delta_{0})=\frac{4\pi^2 P_{0}p^2\tanh\left(\frac{\hbar\omega}{2k_{B}T}\right)}{3\epsilon'}. 
\label{eq:losstan_eq}
\end{eqnarray}
 where  $\tan(\delta_{0})$ expresses the loss tangent in the linear response regime.  In the analysis below we set the hypertangent factor $\tanh\left(\frac{\hbar\omega}{2k_{B}T}\right)$ to unity since the consideration is limited to low temperatures Eq. (\ref{eq:lowbound}).

\section{Master equation formalism for spectral diffusion}
\label{sec:masteq}

\subsection{Spectral diffusion}
\label{sec:sddef}

The {\it spectral diffusion} that is the target of the present work strongly affects microwave absorption by TLSs leading to a  TLS phase decoherence. It is induced by TLS interactions with neighboring ``thermal" TLSs having energies and tunneling amplitudes comparable to the thermal energy. They switch back and forth between their ground and excited states with the quasiperiod $T_{1T} = 1/(\alpha T^3)$, Eq. (\ref{eq:TLSPars}), and modify the energy and detuning of the given TLS.  One can describe the spectral diffusion in terms of the distribution function $W(D, t|D', t')$ for possible detunings $D$ at the time $t$ provided  that $D(t')=D'$. In a three dimensional system with TLS-TLS interaction decreasing with the distance as $1/R^3$, this function has a Lorentzian shape \cite{AndersonKlauder62,BlackHalperin77,LaikhtmanSD86}. In a short time limit $|t-t'| < T_{10} \ll T_{1T}$ (see Eqs. (\ref{eq:upperlim}), (\ref{eq:TLSPars}))  the width $w$ of the Lorentzian distribution,  $\Phi(D-D', t-t')=W(D, t|D', t')$, for the change in detuning, $D-D'$, can be expressed as \cite{AndersonKlauder62,BlackHalperin77,ab15TLSnoise}
\begin{eqnarray}
 w=k_{2}^2|t-t'|, ~
k_{2}^2=\frac{\sqrt{1-x^2}}{T_{sd}^2}, ~ 
\frac{1}{T_{sd}^2}=\frac{\pi^6}{24}\frac{\chi k_{B}T}{\hbar T_{1T}},
\label{eq:TLSDec}
\end{eqnarray}
where the universal dimensionless parameter $\chi\sim 10^{-3}$ is the product of TLS density $P_{0}$ and the average absolute value of TLS-TLS interaction constant  $U_{0}$, $\left<|U(r)|\right>=U_{0}/r^3$ \cite{BlackHalperin77,ab13echo,ab15TLSnoise,YuLeggett88}. The rate $1/T_{sd}$ characterizes the TLS phase decoherence observable in echo experiments.  Spectral diffusion  affects the non-linear absorption if this rate exceeds the TLS relaxation rate $1/T_{10}$ defined in Eq. (\ref{eq:TLSPars}). For a typical microwave field frequency $\omega/(2\pi) \approx 5$GHz, the decoherence stimulated by the spectral diffusion goes faster than that due to the relaxation at reasonably high temperatures $T>T_{l}$ where the temperature $T_{l}$ is given by 
\begin{eqnarray}
T_{l} \approx \left(\frac{12\hbar^7\alpha\omega^6}{\pi^6\chi k_{B}^7}\right)^{1/4} \approx 30  {\rm mK}. 
\label{eq:lowbound}
\end{eqnarray}
Upper (Eq. (\ref{eq:upperlim})) and lower ($T_{l}<T$) constraints on the temperature can be satisfied simultaneously at reasonably  small resonant frequencies $\hbar\omega/k_{B} < 10$K. Our consideration is limited to sufficiently small frequencies and temperatures   $T, \hbar\omega/k_{B} <1$K where the tunneling model \cite{AHV,Ph} is valid, which justifies the relevance of the consideration of the temperature domain restricted by Eqs. (\ref{eq:upperlim}) and  (\ref{eq:lowbound}).   

To describe the loss tangent behavior within the temperature domain $T>T_{l}$ it is natural to attempt to use Eq. (\ref{eq:BlochBas}) with the modified decoherence rate 
\begin{eqnarray}
1/T_{2*}=1/T_{20}+1/T_{sd}. 
\label{eq:modrate}
\end{eqnarray} 
This modifies Eq. (\ref{eq:losstan_eq}) for the TLS loss tangent as \cite{Hunklinger77NonLin,Yu94,ab95DipGap} 
\begin{eqnarray}
\tan(\delta)\approx\frac{\tan(\delta_{0})}{\sqrt{1+\frac{16}{9\pi^2}\Omega_{R0}^2 T_{2*}T_{10}}}. 
\label{eq:losstan_eqMod}
\end{eqnarray}
The accurate treatment of the spectral diffusion described below within the framework of the master equation formalism  predicts that the loss tangent behavior is qualitatively different from  Eq. (\ref{eq:losstan_eqMod}).  

\subsection{Master equations for TLS Bloch vector} 
\label{sec:masteqDer}

The time evolution of the Bloch vector components $d\sigma^{a}= \sigma^{a}(D, t+dt)-\sigma^{a}(D, t)$ can be separated into two parts. The first part described by Eq. (\ref{eq:BlochBas})    includes the quantum mechanical evolution and relaxation    \cite{Yu94,HunklingerReview,Hunklinger77NonLin}. 
The second part accounts for the change in detuning due to spectral diffusion that can be treated classically \cite{BlackHalperin77,LaikhtmanSD86,FaoroIoffe12}.  The change of detuning during the infinitesimal time $dt$ is determined by the conventional Lorentzian probability function $\Phi(D-D', dt)$  introduced above. Evolution of the TLS density matrix in the course of the  spectral diffusion can be described by the master equation as suggested by Laikhtman \cite{LaikhtmanSD86} (see also the textbook \cite{Stratonovichbook})
\begin{eqnarray}
d\pmb{\sigma}_{sd}(D, t)=
\nonumber\\
=\int_{-\infty}^{\infty} dD' \Phi(D-D', dt) 
(\pmb{\sigma}(D', t)-\pmb{\sigma}(D, t)). 
\label{eq:StochEvol}
\end{eqnarray}

This equation is valid if there is no correlation between the TLS evolution before and after the time $t$ \cite{Stratonovichbook}.  Phonon stimulated transitions of resonant  TLSs to the ground state occurs with the quasi-period $T_{10}$ and each transition erases memory about previous TLS  dynamic. Since the time between two transitions of thermal TLS $T_{1T} \approx (\alpha T^3)^{-1}$ is much longer than this quasi-period (see Eqs. (\ref{eq:upperlim}) and (\ref{eq:TLSPars})) each neighboring thermal TLS contributes no more than once to the shift of the resonant TLS energy during the time $T_{10}$. Consequently, there is no correlation of energy shifts during the equilibration time $T_{10}$ which justifies Eq. (\ref{eq:StochEvol}).

Using the Fourier transform representation of the TLS density matrix, Eq. (\ref{eq:lossdefBas}), and the width of the Lorentzian distribution defined in Eq. (\ref{eq:TLSDec}) one can express the time evolution in Eq. (\ref{eq:StochEvol})  as 
\begin{eqnarray}
d\mathbf{f}_{sd}=(e^{-k_{2}^2dt|q|}-1)\mathbf{f}=-k_{2}^2 |q|dt\mathbf{f}. 
\label{eq:FourStoch}
\end{eqnarray}

The complete set of evolution  equations for the density matrix Fourier transforms   (see Eqs. (\ref{eq:losstan_eq})) can be obtained adding Eq. (\ref{eq:FourStoch}) and the Fourier transforms of Eq. (\ref{eq:BlochBas}). Then we get
\begin{eqnarray}
\frac{df^{x}}{dt}=-|q|k_{2}^2f^{x} -i \frac{df^{y}}{dq} -\frac{f^{x}}{2T_{1}}=0,
\nonumber\\
\frac{df^{y}}{dt}=-|q|k_{2}^2f^{y}+i \frac{df^{x}}{dq} -\frac{f^{y}}{2T_{1}} +\Omega_{R}f^{z}=0,
\nonumber\\
\frac{df^{z}}{dt}=-|q|k_{2}^2f^{z}-\Omega_{R}f^{y} -\frac{f^{z}}{T_{1}}+\frac{2\pi \delta(q)}{T_{1}}=0,
\label{eq:BlochMain1}
\end{eqnarray}
where time derivatives should be  set to zero since we are interested in the stationary regime \cite{HunklingerReview}.  All Fourier transforms $f^{a}(q)$ ($a=x, y, z$) should approach $0$ for $q\rightarrow \pm \infty$, which defines the boundary conditions to Eq. (\ref{eq:BlochMain1}). To evaluate the loss tangent, we need to find the Fourier transform $f^{y}(0)$ for $q=0$ in accordance with Eq. (\ref{eq:losstan_eq}). 

Here and in the earlier work \cite{LaikhtmanSD86} the rotating frames approximation assumes the instantaneous adiabatic basis for the Bloch vector following the spectral diffusion of the detuning. This assumption seems to be well justified since a single TLS resonant frequency $\omega_{0}$ is much larger compared to a characteristic rate of the spectral diffusion, however, the collective TLS transitions can be stimulated by their interactions in the course of spectral diffusion  \cite{ab95TLSRelax,ab98book}. This collective dynamics becomes important at very low temperature $T \sim 10$mK \cite{ab13echo}, which is outside of the scope of the present work.

\section{Nonlinear absorption in the presence of spectral diffusion: results and discussion}
\label{sec:Solution}

\subsection{Solution of the master equation for individual TLSs}
\label{sec:masteqIndTLS}

The further analysis of  Eq. (\ref{eq:BlochMain1}) is straightforward but tedious, and its details can be found in the Supplemental Materials, Sec. I \cite{Suppl}. Using the first and third equations, one can express $f^{x}$, $f^{z}$ and $\frac{df^{x}}{dq}$ in terms of $f^{y}$ and obtain the second order differential equation for $f^{y}(q)$ containing delta-function term $\delta(q)$ as a nonhomogeneous part. It is convenient to introduce the new variable $v=(A+k_{2}|q|)^2$ where $A=1/(k_{2}T_{2})$.  Using the substitution  $f^{y}=e^{-\frac{v-A^2}{2}}F(v)$, we get the equation for $F(v)$ in the form similar to the hypergeometric differential equation  (see Ref. \cite{GradshteynRyzhik07,AbramowitzBook}, remember that $v>A^2$)
\begin{eqnarray}
v\frac{d^2 F}{dv^2} - v\frac{d F}{dv}- F\left[\frac{\Omega_{R}^2\eta}{4k_{2}^2}\right]=0, ~ \eta(v)= \frac{1}{1+\frac{A}{\sqrt{v}}}.
\label{eq:KummerFE}
\end{eqnarray}

To account for the $\delta$-function term at  $v=A^2$ ($q=0$, see Eq. (\ref{eq:BlochMain1})) we introduce  the boundary conditions  at $v=A^2$ for the first derivative of the function $F$  in the form \cite{Suppl}
\begin{eqnarray}
\frac{dF(A^2)}{dv}-\frac{1}{2}F(A^2)=-\frac{\pi \Omega_{R}}{2}, 
\label{eq:KummerFB}
\end{eqnarray}
while the second boundary condition is $F(\infty)=0$. 
The parameter of interest $f^{y}(0)$, that determines the loss tangent in Eq. (\ref{eq:lossdefBas}), is equal to $F(A^2)$.

Eqs. (\ref{eq:KummerFE}), (\ref{eq:KummerFB}) represent the most significant results of the present work. They have been solved numerically for this problem and can be extended to other problems of interest. The approximate analytical solution for the TLS loss tangent is derived below in the case of a significant spectral diffusion, $A \ll 1$, and compared to the numerical solution.  
This solution is the subject for comparison   to the experimental data and it helps to understand the non-linear absorption qualitatively.

If the parameter $\eta$ in Eq. (\ref{eq:KummerFE}) can be set to constant then the solution satisfying  the zero boundary condition at infinity is given by the confluent hypergeometric function of the second kind  $F(v)=cU(\eta (\Omega_{R}/k_{2})^2/4, 0, v)$ \cite{AbramowitzBook,Tricomi1947} with the constant $c$ defined by the boundary condition in Eq. (\ref{eq:KummerFB}). Comparing the first term in Eq. (\ref{eq:KummerFE})  with the second and third terms one can see that the solution of Eq. (\ref{eq:KummerFE})  should change remarkably compared to its maximum $F(A^2)$ ($q=0$)  at $v \sim v_{*} =$ min$(1, k_{2}^2/\Omega_{R}^2)$. 
Assuming that $v_{*}$ determines the typical value of the parameter  $\eta$ one can expect that $\eta \approx 1$ for $v_{*} \gg A^2$, while in the opposite case $\eta \approx 1/2$. This is confirmed by the comparison of analytical and numerical solutions \cite{Suppl}.

In either case one can evaluate the parameter of interest $f^{y}(0)=F(A^2)$ (see Eq.  (\ref{eq:lossdefBas}))  using the boundary condition Eq. (\ref{eq:KummerFB}) and the identity $dU(a, b, v)/dv=-aU(a+1, b+1, v)$ \cite{AbramowitzBook}  as
\begin{eqnarray}
f^{y}(0)=\frac{\pi\Omega_{R}}{1+\frac{\eta B^2}{2}\frac{U(\eta B^2/4, 0, A^2)}{U(1+\eta B^2/4, 1, A^2)}}, ~ B=\frac{\Omega_{R}}{k_{2}}.
\label{eq:KummerSol3}
\end{eqnarray}

Consider the case $v_{*} > A^2$ ($\Omega_{R} < k_{2}^2 T_{1}$ and we set $\eta \approx 1$). 
In this case one can use the asymptotic behaviors of confluent hypergeometric functions  $U(a, 0, v)\approx \Gamma(1+a)^{-1}$, $U(1+a, 1, v) \approx (\ln(1/v)-2\gamma-\psi(1+a))\Gamma(1+a)^{-1}$ and approximate the digamma function as $\psi(1+a)=\psi(1+B^2/4) \approx 2\ln(B/2+e^{-\gamma/2})$, where $\gamma\approx 0.5772$ is the Euler constant. Then Eq. (\ref{eq:KummerSol3}) can be represented as 
\begin{eqnarray}
f^{y}(0)=\frac{\pi \Omega_{R}}{1+\frac{\Omega_{R}^2}{k_{2}^2}\ln\left(\frac{2e^{-\gamma}k_{2}^2T_{2}}{2e^{-\frac{\gamma}{2}}k_{2}+\Omega_{R}}\right)}. 
\label{eq:ANSintAsymp} 
\end{eqnarray}
This solution is in excellent  agreement with the exact numerical solution 
if the argument of the logarithm exceeds $2$ \cite{Suppl}. 

In the opposite  limit of a very large field $B > 1/A$ ($\Omega_{R} > k_{2}^2 T_{1}$ and we set $\eta \approx 1/2$) one can approximate Eq. (\ref{eq:KummerSol3}) as \cite{Suppl}
\begin{eqnarray}
f^{y}(0) \approx \frac{\pi\Omega_{R}}{\sqrt{1+\frac{B^2}{2A^2}}}.
\label{eq:KummerSol3a}
\end{eqnarray}
This result is identical to the stationary solution of  Eq. (\ref{eq:BlochBas}) or  Eq. (\ref{eq:BlochMain1}) at $k_{2}=0$ as given by Eq. (\ref{eq:KummerSol3a0}). It 
does not depend on the spectral diffusion. The spectral diffusion is not significant in this regime because the typical detunings of resonant TLSs  contributing to absorption, $D \sim  \Omega_{R}$ \cite{FaoroIoffe15}, exceed a TLS frequency shift $D_{sd} \sim k_{2}^2 T_{1}$ due to the spectral diffusion occurring during the  time $T_{1}$ separating two resonant TLS relaxation events (see Eq. (\ref{eq:TLSDec})). Substituting  Eq. (\ref{eq:KummerSol3a}) into the loss tangent definition, Eq. (\ref{eq:lossdefBas}) and performing the integration one obtains the earlier established  behavior of the loss tangent Eq. (\ref{eq:losstan_eq}), \cite{PhillipsReview,HunklingerReview}.

The solution of time dependent master equations has been obtained in Ref. \cite{LaikhtmanSD86} ignoring relaxation terms (i. e. setting $T_{1}=T_{2}=\infty$) for a finite pulse duration. 
Although it has a certain similarity with the present work it leads to a full suppression of absorption in the infinite duration time limit that is the consequence of the lack of  dissipation needed for the correct description of absorption in a steady state regime. 

\subsection{Evaluation of the loss tangent}
\label{sec:lt}

\begin{figure}[h!]
\centering
\includegraphics[width=\columnwidth]{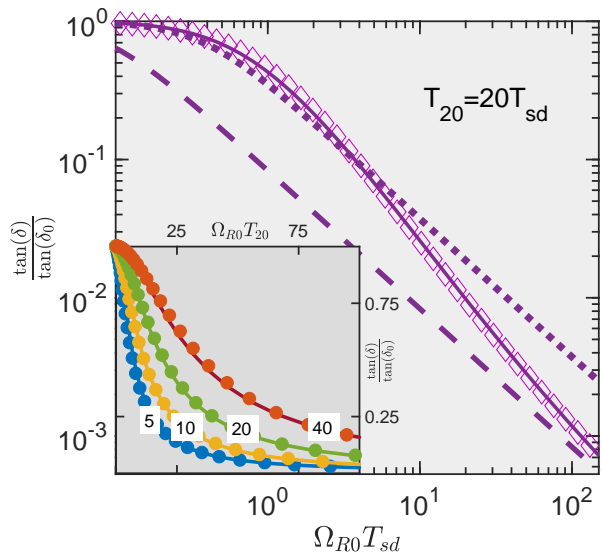}
\caption{ The loss tangent vs Rabi frequency in the presence of the fast spectral diffusion $T_{20} \gg T_{sd}$.
The main graph shows the numerical solution (solid line), analytical interpolation of Eq. (\ref{eq:losstan_ANS}) (diamonds), the solution, Eq. (\ref{eq:losstan_eq}), ignoring the spectral diffusion (dashed line), and the solution Eq. (\ref{eq:losstan_eqMod}) 
with the modified decoherence rate  (dotted line) in the case $T_{20}=2T_{10}=20T_{sd}$. In the inset the numerical (solid lines) and analytical  (Eq. (\ref{eq:losstan_ANS}), circles) solutions are compared for different relative rates of  spectral diffusion $T_{20}/T_{sd}$ (indicated at each line).} 
\label{fig:Fig2NonLinAbs}
\end{figure}

The loss tangent is defined by the integral in Eq. (\ref{eq:lossdefBas}) over different tunneling amplitudes and dipole moment orientations of contributing two-level systems. It has been evaluated using the exact numerical solution of Eq. (\ref{eq:KummerFE}) as shown in Fig. \ref{fig:Fig2NonLinAbs} by solid lines for several ratios of spectral diffusion and relaxation rates  ($T_{20}/T_{sd}$). The predicted behaviors differ from that for the loss tangent in the absence of spectral diffusion, Eq. (\ref{eq:losstan_eq}), as shown in Fig. \ref{fig:Fig2NonLinAbs} by dashed lines  and from the ``corrected" Eq. (\ref{eq:losstan_eq}) with the modified decoherence time $T_{2*}$ ($1/T_{2*}=1/T_{20}+1/T_{sd}$) as shown in Fig. \ref{fig:Fig2NonLinAbs} by the dotted line. 

Below, we derive the analytical interpolation for the loss tangent   suitable for the analysis of experimental data in the case of a significant spectral diffusion $T_{sd} < T_{10}$; otherwise, the loss tangent is always determined by Eq. (\ref{eq:losstan_eq}).  In the case of an intermediate Rabi frequency, $1/T_{sd}<\Omega_{R0}<T_{10}/T_{sd}^2$, one can use the approximate solution, Eq. (\ref{eq:ANSintAsymp}), substituted into the integral in Eq. (\ref{eq:lossdefBas}). Then, one can evaluate this integral with  logarithmic accuracy  as  $\tan(\delta_{0})\ln(1+3l)/(\Omega_{R0}T_{sd})^2$ with the logarithmic factor $l=\ln(\Omega_{R0}T_{sd})/\ln(D_{sd}/\Omega_{R0})$ \cite{Suppl}. 
In the opposite case of large Rabi frequency, $T_{10}/T_{sd}^2<\Omega_{R0}$, the loss tangent is determined by Eq. (\ref{eq:losstan_eq}). One can combine these two asymptotic behaviors within the single interpolation formula 
\begin{eqnarray}
\tan(\delta_{sd})=\frac{\tan(\delta_{0})}{1+
\frac{\Omega_{R0}^2T_{sd}^2}{\frac{3\pi\Omega_{R0}T_{sd}^2}{4\sqrt{2}T_{20}}+\ln(1+3l_{*})}}
\label{eq:losstan_ANS}
\end{eqnarray}
with the logarithmic factor $l_{*}$ defined as
\begin{eqnarray}
 l_{*}=\frac{\ln(d_{1}+c_{1}\Omega_{R0}T_{sd})}{\ln\left(d_{2}+\frac{c_{2}T_{20}}{\Omega_{R0}T_{sd}^2}\right)}.
\label{eq:losstan_ANSl_*}
\end{eqnarray}
The numerical constants $c_{1}, d_{1}, c_{2}, d_{2} \sim 1$  in the definition of $l_{*}$ were introduced to describe the crossover between two regimes.  They were estimated as $0.5, 3, 0.55, 1.2$, respectively, by fitting the exact numerical solution with Eq. (\ref{eq:losstan_ANS})  \cite{Suppl}. As shown in Fig. \ref{fig:Fig2NonLinAbs} this fit is perfectly consistent with the numerical solution and, therefore, it can be used to analyze experimental data. 
In the asymptotic regime of a large external field, $T_{10}/T_{sd}^2 < \Omega_{R0}$, one can ignore the factor $\ln(1+3l_{*})$ in  the denominator in Eq. (\ref{eq:losstan_ANS}), which leads to  Eq. (\ref{eq:losstan_eq}), while in the opposite regime, the logarithmic term $l_{*}\approx l$ dominates in the denominator. 

Using Eq. (\ref{eq:losstan_ANSl_*}) one can estimate the crossover Rabi frequencies separating linear and non-linear regimes ($\Omega_{c1}$) and regimes of significant and negligible spectral diffusion ($\Omega_{c2}$) characterized by the loss tangent behaviors $\tan(\delta) \propto \Omega_{R0}^{-2}$ and $\tan(\delta) \propto \Omega_{R0}^{-1}$, respectively. We assume that the spectral diffusion is much faster than the relaxation, i. e. $T_{sd} \ll T_{20}$. For the first crossover representing the non-linear threshold one can estimate $\ln(1+3l_{*}) \approx 3\ln(d_{1})/\ln(T_{20}/T_{sd})$ within the logarithmic accuracy. The threshold Rabi frequency  where the loss tangent gets smaller than its linear response theory value by the factor of two can be expressed as 
\begin{eqnarray}
\Omega_{c1} \sim \frac{3.3}{T_{sd}\ln\left(\frac{T_{20}}{T_{sd}}\right)}.
\label{eq:OmcHT0}
\end{eqnarray} 
This crossover frequency is proportional to the squared temperature. Consequently, the non-linear threshold intensity increases as the fourth power of the temperature. 

The second crossover frequency can be estimated setting two factors in the denominator of Eq. (\ref{eq:ANSintAsymp}) responsible for the spectral diffusion ($\ln(1+3l_{*})$) and relaxation ($3\pi\Omega_{R0}T_{sd}^2/(4\sqrt{2}T_{20})$) regimes equal to each other. Then within the logarithmic accuracy one gets
\begin{eqnarray}
\Omega_{c2} \sim  \frac{0.6 T_{20}}{T_{sd}^2}\ln\left(3\ln\left(\frac{T_{20}}{T_{sd}}\right)\right).
\label{eq:OmcLT0}
\end{eqnarray} 

The estimates for the threshold Rabi frequency separating linear and non-linear regimes has been obtained in Ref. \cite{LaikhtmanSD86} using qualitative arguments and in Ref. \cite{Galperin88} using the rigorous perturbation theory analysis. Both estimates are consistent with the present work; moreover the estimate of Ref. \cite{Galperin88} differes only by the factor of $\pi/3.3 \approx 0.95$, which is an excellent agreement. 
The loss tangent intensity dependence predicted by Eq. (\ref{eq:losstan_ANS}) is also consistent with the qualitative estimates of Refs. \cite{LaikhtmanSD86,Galperin88}, yet here it is obtained in a rigorous form. The weakening of that dependence at higher intensities to the inverse square root behavior to our knowledge was not considered before the present work. 

\subsection{Loss tangent behavior in regimes where the present theory is not applicable}
\label{sec:ltoutscope}

The analytical interpolation for the loss tangent given by Eq. (\ref{eq:losstan_ANSl_*}) is applicable only if the spectral diffusion  is faster then  the TLS relaxation  
($T_{sd}<T_{10}$) that takes place at sufficiently large temperature $T>T_{l}$, Eq. (\ref{eq:lowbound}), 
and the resonant TLS relaxation is faster than that of thermal TLSs responsible for the spectral diffusion suggesting  $k_{B}T < \hbar\omega$, Eq. (\ref{eq:upperlim}). For the typical microwave experimental frequency of $5$GHz this limits the theory applicability to temperatures $30$mK $<T< 200$mK \cite{ab13LZTh}. Below we consider the loss tangent outside of this temperature domain. 

At low temperatures $T_{sd}<T_{10}$ Eq. (\ref{eq:BlochMain1})  is still applicable. In this regime  its numerical solution shown in Supplementary Materials \cite{Suppl} is almost identical to Eq. (\ref{eq:lossdefBas}) 
so that the non-linear absorption can be very well described by Eq. (\ref{eq:losstan_eq}) ignoring the spectral diffusion. It is noticeable that in the crossover regime of $T_{sd}=T_{10}$  the numerical solution, analytical interpolation оф Eq. (\ref{eq:losstan_ANSl_*}) and standard model described by Eq. (\ref{eq:losstan_eq}) predict almost identical behaviors. At higher temperatures ($T_{sd}<T_{10}$) one can use Eq. (\ref{eq:losstan_ANSl_*}) until $k_{B}T < \hbar\omega$. 

At high temperature, $k_{B}T > \hbar\omega$, the master equation formalism is no longer applicable and we cannot obtain the accurate quantitative solution for the loss tangent. Below, we suggest  the qualitative arguments to predict the non-linear absorption behavior in this regime ignoring possible logarithmic dependencies. 

The crossover Rabi frequency $\Omega_{c}$ separating linear and non-linear regimes can be estimated considering the probability of absorption during the resonant TLS relaxation time $T_{10}>T_{1T}$ (remember that $T_{1T}=1/(\alpha T^3)$ estimates  the relaxation time of thermal TLSs responsible for the spectral diffusion). During that time the TLS energy passes through the resonance $T_{10}/T_{1T}$ times and the probability of absorption during each passage is given by $\Omega_{R}^2T_{sd}^2$ so the total absorption probability during the time $T_{10}$ can be estimated as $P_{abs} \approx \Omega_{R}^2T_{sd}^2 T_{10}/T_{1T}$, provided that $P_{abs} <1$. The saturation in absorption is expected to take place at $P_{abs} \sim 1$, suggesting the crossover Rabi frequency 
\begin{eqnarray}
\Omega_{c1} \sim \frac{1}{T_{sd}}\sqrt{\frac{T_{1T}}{T_{10}}}.
\label{eq:OmcHT}
\end{eqnarray} 
The more accurate estimate of the non-linear threshold for $\hbar\omega_{0} <k_{B}T$ including numerical and logarithmic factors can be found in Ref. \cite{Galperin88}. Using definitions of Eqs. (\ref{eq:TLSPars}) and (\ref{eq:TLSDec})  one can estimate the temperature dependence of the threshold Rabi frequency as $\Omega_{c1} \propto T$, while the threshold intensity behaves as $I_{c} \propto \Omega_{c1}^2 \propto T^2$. 

At small Rabi frequency $\Omega_{R} < \Omega_{c}$ the loss tangent can be described by the linear response theory expression given by Eq. (\ref{eq:losstan_eq}), while at higher Rabi frequencies the inverse intensity dependence $\tan(\delta) = \tan(\delta_{0})(\Omega_{c}/\Omega_{R})^2$ is expected  similarly to the intermediate temperature regime and in agreement with Ref. \cite{Galperin88}. This dependence holds until the Rabi frequency is smaller than the typical maximum spectral diffusion range ($\Omega_{R} < \Omega_{c2} \sim T_{1T}/T_{sd}^2$) while at larger intensities the nonlinear absorption is no more sensitive to the spectral diffusion and can be described by Eq. (\ref{eq:losstan_eq}). It is straightforward to check that different asymptotic behaviors are consistent with each other at the crossovers. 

In the case of thermal energy exceeding the field quantization energy there exists the emergence of relaxational absorption that contributes to the loss tangent as $\tan(\delta_{rel}) \approx \tan(\delta_{0})\alpha T^3/\omega$ for $\hbar\Omega_{R0}<k_{B}T$ \cite{HunklingerReview}. One should notice that the relaxational absorption is suppressed exponentially in the opposite limit of low temperatures given by Eq. (\ref{eq:upperlim})  so it can be neglected there.  Here it limits the maximum temperature to $10\hbar\omega/k_{B}$ to keep the crossover estimates valid and limits the maximum Rabi frequency to $\omega (\hbar\omega/(k_{B}T))^3$ to keep the resonant absorption dominating. We assume both constraints to be satisfied as indicated in the table \ref{tbl:summary}.

\subsection{Discussion of experiment}
\label{sec:exp}

The theoretical predictions for the nonlinear absorption in the presence of spectral diffusion was made in Ref. \cite{Hunklinger77NonLin} using Eq. (\ref{eq:losstan_eqMod}) with a modified decoherence rate as described in Eq. (\ref{eq:modrate}).  
These predictions do  not have any specific domain of relevance, yet they do not deviate dramatically from the predictions of the present theory (see Eq. (\ref{eq:losstan_eqMod}) and the dotted line in Fig. \ref{fig:Fig2NonLinAbs} calculated at spectral diffusion rate exceeding the TLS relaxation rate by the factor of $10$ similarly to Ref. \cite{Hunklinger77NonLin}). Particularly, this approximate agreement could be the reason for the conclusion of Ref. \cite{Hunklinger77NonLin} about the relevance of that approach to the experiments. 

One should notice that the temperature dependence of the threshold intensity, that separates linear and non-linear regimes, is experimentally observed as $I\propto T^4$, which conflicts with the theoretical model of Ref. \cite{Hunklinger77NonLin} predicting $I\propto T^2$. However, it is consistent with the present theory determining the crossover Rabi frequency as the inverse TLS decoherence rate, $1/T_{sd} \propto T^{2}$, Eq. (\ref{eq:TLSDec}). The critical intensity is determined by the squared Rabi frequency, leading to the $T^{4}$ dependence in agreement with the experiment \cite{Hunklinger77NonLin} (cf. Ref. \cite{Golding83}).  

We hope that the present work provides solid background for future experiments that can verify the predictions of the theory. The analytical interpolation of Eq. (\ref{eq:ANSintAsymp}) should serve as a guideline for the data analysis since it covers not only asymptotic regimes but crossovers between them. It is important that the novel behavior takes place in the domain of Rabi frequencies (intensities) restricted from both lower and upper sides (see Eqs. (\ref{eq:OmcLT0}) and (\ref{eq:OmcLT0})) and therefore the ratio of relaxation and decoherence times $T_{10}/T_{sd}$ determining 
the size of this domain should be chosen sufficiently large.  The other significant problem of a finite pulse duration can also affect the experimental data \cite{Golding83,LaikhtmanSD86}. This duration should exceed the TLS relaxation time to make the theory applicable \cite{LaikhtmanSD86}.

The numerous measurements of the non-linear loss tangent has been performed in Josephson junction qubits and resonators (see e. g. Refs. \cite{Martinis05,PaikOsborn10,UstinovScience12,LisenfeldReview17}). All of them use Eqs. (\ref{eq:losstan_eq}) or (\ref{eq:losstan_eqMod}) as a guideline for the experimental data analysis in a broad temperature range including the temperatures where the spectral diffusion is significant ($30$mK $<T< 200$mK). In the present work we demonstrate that these equations are not applicable in this regime and should be modified according to Eq. (\ref{eq:ANSintAsymp}). The experimental observations, indeed, show the intensity dependence different from $1/\sqrt{I}$ (Eqs. (\ref{eq:losstan_eq}), (\ref{eq:losstan_eqMod})) in the non-linear regime. However, they show weakening compared to this dependence while the present work predicts its strengthening. In our opinion this  discrepancy can be understood, assuming that the standard model of interacting TLSs in three dimensions is not quite relevant for quantum two level systems in Josephson junctions. Particularly,   amorphous films used in Josephson junctions possess the reduced dimensionality that can affect the interaction between TLSs and/or their statistics \cite{Osborn132TLS,ab18MoshePeter}. The theory should be modified accordingly, which is beyond the scope of the present work.

\section{Conclusion}
\label{sec:concl}



\begin{table*}
 \caption{Summary of intensity dependencies of the TLS loss tangent at different temperatures below $1$K where TLS model is applicable. The intermediate intensity regime of significant spectral diffusion is not available at low temperatures where the relaxation is faster than the phase decoherence as indicated by ``n. a.'' notartion in a corresponding cell of the table. TLS parameters used within the table are introduced in Eqs. (\ref{eq:TLSPars}), (\ref{eq:TLSDec}) while the results are given in Sec. \ref{sec:lt}. The restrictions for the maximum Rabi frequency and temperature are introduced to avoid the emergence of relaxational absorption (see the end of Sec. \ref{sec:ltoutscope}) and keep the rotating frames approximation valid.} 
 \label{tbl:summary}
\begin{tabular}{|c|c|c|c|c|c|}
\hline
\multirow{2}{*}{} & \multirow{2}{*}{$\Omega_{c1}$} & \multirow{2}{*}{$\Omega_{c2}$} & \multicolumn{3}{c|}{$\frac{\tan(\delta)}{\tan(\delta_{0})}$} \\
\cline{4-6}
  & & &$\Omega_{R0}<\Omega_{c1}$   & $\Omega_{c1}<\Omega_{R0}<\Omega_{c2}$ & $\Omega_{c2}<\Omega_{R0}< \omega$ min$\left(1, \left(\frac{\hbar\omega}{k_{B}T}\right)^3\right)$ 
 \\
\cline{1-6}
 $T< \left(\frac{12\hbar^7\alpha\omega^6}{\pi^6\chi k_{B}^7}\right)^{1/4}$& $\frac{1.67}{T_{20}}$ & $\frac{1.67}{T_{20}}$ &  $1$ & n. a. & $\frac{1.67}{\Omega_{R0}T_{20}}$\\
\hline
\cline{1-6}
$\left(\frac{12\hbar^7\alpha\omega^6}{\pi^6\chi k_{B}^7}\right)^{1/4}<T<\frac{\hbar\omega}{k_{B}}$  & $\frac{3.3}{T_{sd}\ln\left(\frac{T_{20}}{T_{sd}}\right)}$ & $\frac{0.6 T_{20}}{T_{sd}^2}\ln\left(3\ln\left(\frac{T_{20}}{T_{sd}}\right)\right)$ & $1$ & $\frac{\ln(1+3\ln(\Omega_{R0}T_{sd}))}{\Omega_{R0}^2T_{sd}^2\ln\left(\frac{T_{20}}{\Omega_{R0}T_{sd}}\right)}$ & $\frac{1.67}{\Omega_{R0}T_{20}}$ \\
\hline
\cline{1-6}
$\frac{\hbar\omega}{k_{B}}<T<$ min($1$K, 10$\hbar\omega$) & $\frac{1}{T_{sd}}\sqrt{\frac{T_{1T}}{T_{10}}}$ & $\frac{T_{1T}}{T_{sd}^2}$ & $1$ & $\frac{T_{1T}^2}{\Omega_{R0}^2T_{sd}^4}$  & $\frac{1.67}{\Omega_{R0}T_{20}}$ \\
\hline
\end{tabular}

\end{table*}

The present work suggests the resolution of the long-standing problem of the non-linear absorption by interacting two level systems (TLSs) in low temperature amorphous solids. The solution to this problem is obtained using the master equation formalism developed for the spectral diffusion of TLS resonant frequencies induced by their long-range interactions. It is demonstrated that the spectral diffusion extends the domain of a linear absorption compared to the non-interacting model, Eq. (\ref{eq:losstan_eq}), to Rabi frequencies  of the order of the TLS phase decoherence rate ($\Omega_{R0} \sim 1/T_{sd}$, see Eq. (\ref{eq:losstan_ANS})) provided that  this rate exceeds the relaxation rate $1/T_{10}$. At larger Rabi frequencies, $1/T_{sd}<\Omega_{R0}<T_{20}/T_{sd}^2$, the loss tangent decreases inversely proportionally to the intensity of the external field ($\Omega_{R0}^{-2}$, see Eq. (\ref{eq:losstan_ANS})). {\it This new behavior can be understood assuming that the absorption by almost all TLSs, passing the resonance due to the spectral diffusion, is saturated for slow passage $\Omega_{R0}T_{sd} >1$} (cf. Ref. \cite{ab13LZTh,ab14LZExp} where the slow resonance passage due to a bias sweep leads to the similar behavior). {\it TLSs passing the resonance absorb nearly the same energy $\hbar\omega$, while the absorption by other TLSs is negligible. Consequently, the absorbed energy is weakly sensitive to the field intensity and the loss tangent is inversely proportional to that intensity.}
At larger intensities, the resonant domain size $\Omega_{R}$ exceeds the spectral diffusion range   $T_{20}/T_{sd}^2$ \cite{FaoroIoffe12} and its increase with the intensity restores the earlier predicted behavior $\tan(\delta)\propto \Omega_{R0}^{-1}$ , Eq. (\ref{eq:losstan_eq}). For the typical microwave frequency $\omega/(2\pi) \sim 5$GHz the spectral diffusion remains significant ($T_{10}/T_{sd} > 1$) at temperatures exceeding $30$mK (see Eqs. (\ref{eq:TLSPars}), (\ref{eq:TLSDec}) and Ref. \cite{ab13LZTh}). The results of the consideration are summarized in the table \ref{tbl:summary}. 

The theory is fully extendable to the non-linear absorption of acoustic waves. The non-linear internal friction can be described replacing the Rabi frequency in Eq. (\ref{eq:losstan_ANS}) with the product of the TLS-phonon  interaction constant and the strain field ($\gamma\epsilon$, see reviews \cite{HunklingerReview,PhillipsReview}). 
Although the present theory is not directly applicable to high temperatures $T>\hbar\omega/k_{B}$ where the thermal energy exceeds the external field quantization energy, the similar behavior of the non-linear absorption is expected in this regime as well with modified crossover intensity as described in Sec. \ref{sec:ltoutscope} and Table \ref{tbl:summary}.  


The predictions of our theory can be verified experimentally by measuring the non-linear absorption of acoustic or electromagnetic waves in ``ordinary'' glasses where the TLS model \cite{AHV,Ph,BlackHalperin77} is relevant. The analytical expression for the loss tangent  given by Eq. (\ref{eq:ANSintAsymp}) can be used as a guide line for the data analysis in a broad domain of parameters (see Table \ref{tbl:summary}). 
The results of previous measurements seem to be inconclusive (see discussion in Sec. \ref{sec:exp}) because the range of field intensities is insufficiently broad to distinguish between different intensity dependencies. According to our analysis  the conclusive measurements can be performed varying the external field intensity by one or two orders of magnitude in the non-linear regime.

The predicted strengthening of the loss tangent intensity dependence in the non-linear regime ($\propto I^{-1}$, Eq. (\ref{eq:losstan_ANS})) contrasts with the observed weakening of the intensity dependence of microwave absorption in Josephson junction qubits  \cite{PaikOsborn10,Osborn11,Osborn132TLS,ab17Katz,LisenfeldReview17,FaoroIoffe12}. The generalization of the present theory to  low dimensions and modified TLS distribution compared to the standard tunneling model are possibly needed to interpret those experiments.


\begin{acknowledgments}
This research is partially supported by the National Science Foundation (CHE-1462075)  and the Tulane University Carol Lavin Bernick Faculty Grant.
 Authors acknowledge stimulating discussions with Moshe Schechter, Alexander Shnirman, George Weiss, Christian Enss, Kevin Osborn, Joseph Popejoy and Ma'ayan Schmidt.  
\end{acknowledgments}

\bibliography{MBL}

\newpage

\begin{widetext}

{\Huge Supplemental Materials}

\section{Derivation of the equation and boundary condition (Eqs. 14, 15 in the main text) for the function $F$}

According to the main text, Eq. 4 there, the loss tangent is determined by the stationary solution for the Fourier transform  $f^{y}(0)$ of the TLS Bloch vector $y$ component taken with respect to its detuning. The vector  $\mathbf{f}(q)$ of Bloch vector's Fourier transforms  obeys the equations (Eq. 13 in the main text)
\begin{eqnarray}
-|q|k_{2}^2f^{x} -i \frac{df^{y}}{dq} -\frac{f^{x}}{2T_{1}}=0,
\nonumber\\
-|q|k_{2}^2f^{y}+i \frac{df^{x}}{dq} -\frac{f^{y}}{2T_{1}} +\Omega_{R}f^{z}=0,
\nonumber\\
-|q|k_{2}^2f^{z}-\Omega_{R}f^{y} -\frac{f^{z}}{T_{1}}+\frac{2\pi \delta(q)}{T_{1}}=0,
\label{eq:smBlochMain1}
\end{eqnarray}
Here we use Eq. (\ref{eq:smBlochMain1}) to derive Eqs. 14 and 15 from the main text.

Consider Eq. (\ref{eq:smBlochMain1}). Using the first and third equations, one can express $f^{x}$, $f^{z}$ and $\frac{df^{x}}{dq}$ in terms of $f^{y}$ as 
 \begin{eqnarray}
 f^{x}=-\frac{i \frac{df^{y}}{dq}}{|q|k_{2}^2+\frac{1}{2T_{1}}}, ~
 f^{z}=-\frac{\Omega_{R}f^{y}}{|q|k_{2}^2+\frac{1}{T_{1}}}+2\pi \delta(q), ~
 \frac{df^{x}}{dq} =-\frac{i \frac{d^2f^{y}}{dq^2}}{|q|k_{2}^2+\frac{1}{2T_{1}}}+
 \frac{i {\rm sign}(q)k_{2}^2\frac{df^{y}}{dq}}{(|q|k_{2}^2+\frac{1}{2T_{1}})^2}.
\label{eq:smBlochMain2}
\end{eqnarray}
Substituting these results into the second equation in Eq. (\ref{eq:smBlochMain1}), we got 
\begin{eqnarray}
\left(\frac{1}{2T_{1}}+|q|k_{2}^2\right)f^{y}-\frac{\frac{d^2f^{y}}{dq^2}}{|q|k_{2}^2+\frac{1}{2T_{1}}}+
\frac{{\rm sign}(q)k_{2}^2\frac{df^{y}}{dq}}{(|q|k_{2}^2+\frac{1}{2T_{1}})^2}+ 
\frac{\Omega_{R}^2f^{y}}{|q|k_{2}^2+\frac{1}{T_{1}}}
-2\pi \Omega_{R}\delta(q)=0. 
\label{eq:smBlochMain3}
\end{eqnarray}
Introducing the dimensionless variable 
\begin{eqnarray}
Q=k_{2}q 
\label{eq:smQ}
\end{eqnarray}
and parameters 
\begin{eqnarray}
A=\frac{1}{2k_{2}T_{1}}=\frac{1}{k_{2}T_{2}}, ~ B=\frac{\Omega_{R}}{k_{2}}, 
\label{eq:smPars}
\end{eqnarray}
we get the equation for $f_{y}$ in the form 
\begin{eqnarray}
\frac{d^2f^{y}}{dQ^2}-\frac{df^{y}}{dQ}\frac{{\rm sign}(Q)}{A+|Q|} -
f^{y}\left((A+Q)^2+\frac{B^2}{2}\frac{A+|Q|}{A+|Q|/2}\right)+2\pi\Omega_{R}A\delta(Q)=0.
\label{eq:smBlochSigmay}
\end{eqnarray}
This equation is symmetric with respect to the replacement $Q \rightarrow -Q$, so the solution should be the even function of $Q$. 
Consequently, the $\delta$-function term can be treated as a boundary condition for the derivative of $f^{y}(Q)$ at $Q=0$ in the form $df^{y}/dQ=-\pi A\Omega_{R}$ in  Eq. (\ref{eq:smBlochSigmay}) considered for $Q>0$. The second ``common sense" boundary condition is $f^{y}(\infty)=0$. 

Then it is convenient to introduce the new variable $v=(A+|Q|)^2$ and separate the exponential dependence in the function $f^{y}$ representing it as $f^{y}=e^{-\frac{v-A^2}{2}}F(v)$. The function $F(v)$ satisfies the equation 
\begin{eqnarray}
v\frac{d^2 F}{dv^2} - v\frac{d F}{dv}- F\left[\frac{B^2\eta}{4}\right]=0, ~ \eta(v)= \frac{1}{1+\frac{A}{\sqrt{v}}},
\label{eq:smKummerFE}
\end{eqnarray}
while the boundary condition reformulated at $v=A^2$ can be rewritten in the form 
\begin{eqnarray}
\frac{dF(A^2)}{dv}-\frac{1}{2}F(A^2)=-\frac{\pi \Omega_{R}}{2}.
\label{eq:smKummerFB}
\end{eqnarray}
The second boundary condition takes the standard  form $F(\infty)=0$. Eq. (\ref{eq:smKummerFE}) together with the boundary condition given by  Eq. (\ref{eq:smKummerFB}) are used within the main text (Eqs. 14 and 15 there). 

\section{Approximate analytical solution for the absorption rate $\Omega_{R}f^{y}(0)$ and its comparison to the numerical solution}

Consider Eq. (\ref{eq:smKummerFE}). The parameter $\eta$ in this equation changes between $1/2$ at $v-A^2 \ll A^2$ and $1$ for $v \gg A^2$. 
It is quite natural to expect that in different asymptotic regimes one can replace $\eta$ with $1$, if significant values of the variable $v$ exceed $A^2$, or with $1/2$ otherwise. To address this problem one can seek for the solution of Eq. (\ref{eq:smKummerFE}) in the form 
\begin{eqnarray}
F(V)=ce^{-\int_{A^2}^{x}p(x) dx}, ~ c=\frac{\pi \Omega_{R}}{1+2p(A^2)}, 
\label{eq:smpsubs}
\end{eqnarray} 
where the constant $c$ is chosen 
to satisfy the boundary condition, Eq. (\ref{eq:smKummerFB}). This constant $c$ is equal to $F(A^2)$ which is the target of the present consideration. To find $c$ one should calculate $p(A^2)$. Substituting Eq. (\ref{eq:smpsubs}) into Eq. (\ref{eq:smKummerFE}) we obtain the Riccati equation for the function $p$ in the form 
\begin{eqnarray}
\frac{dp}{dv}=p(1+p)-\frac{B^2}{4(v+A\sqrt{v})}. 
\label{eq:smKummerFE5}
\end{eqnarray}
In the limit of $v\rightarrow \infty$ two possible limiting behaviors of the function $p$ can be expected, including $p \rightarrow 0$ or $p\rightarrow -1$. In the latter case the function $F$ approaches infinity exponentially for $v\rightarrow \infty$ (see Eq. (\ref{eq:smpsubs})), so it is not acceptable. Consequently, the solution of interest should satisfy the boundary condition $p(\infty)=0$.   

The formal solution of Eq. (\ref{eq:smKummerFE5}) can be written as 
\begin{eqnarray}
p(v)=Ce^{v+\int_{0}^{v}p(w)dw}+\frac{B^2}{4}\int_{v}^{\infty}dw\frac{e^{v-w-\int_{v}^{w}dw'p(w')}}{w+A\sqrt{w}},
\label{eq:smKummerFE6}
\end{eqnarray}
where $C$ is an arbitrary constant. This solution approaches zero at large $v$ only if $C=0$. Otherwise, at a finite $C$ the second term in the right hand side of Eq. (\ref{eq:smKummerFE6}) approaches zero for $p(\infty)=0$, while the first term remains finite, which conflicts with the boundary condition $p(\infty) = 0$. Consequently, Eq. (\ref{eq:smKummerFE6}) takes the form 
\begin{eqnarray}
p(v)=\frac{B^2}{4}\int_{v}^{\infty}dw\frac{e^{v-w-\int_{v}^{w}dw'p(w')}}{w+A\sqrt{w}}. 
\label{eq:smKummerFE7}
\end{eqnarray}
In the case of interest $v=A^2$ this equation can be rewritten as 
\begin{eqnarray}
p(A^2)=\frac{B^2}{4}\int_{A^2}^{\infty}dw\frac{e^{A^2-w-\int_{A^2}^{w}dw'p(w')}}{w+A\sqrt{w}}. 
\label{eq:smKummerFE8}
\end{eqnarray}

The integral in Eq. (\ref{eq:smKummerFE8}) is logarithmic if the absolute value of the negative exponent $-A^2+w+\int_{A^2}^{w}dw'p(w') \sim (w-A^2)(1+p(A^2))$ approaches $1$ for $w \gg A^2$. This is true if this exponent is much less then $1$ for $w\sim A^2$. In the latter case one can estimate that exponent to be $A^2(1+p(A^2))$. Assuming that it is much less then $1$ and $p(v)$ is a smooth function,  one can estimate $p(A^2)$ by evaluating the logarithmic integral in Eq. (\ref{eq:smKummerFE8}) as
\begin{eqnarray}
p(A^2)=\frac{B^2}{4}\int_{A^2}^{\infty}dw\frac{e^{-(w-A^2)(1+p(A^2))}}{w+A\sqrt{w}}\approx B^2 \ln\left(\frac{1}{(1+p(A^2))A^2}\right).
\label{eq:smKummerFE9}
\end{eqnarray}
Consequently, the assumption $A^2(1+p(A^2)) \ll 1$ is satisfied if  $A \ll 1$ and $AB \ll 1$.  Then, the integral in Eq. (\ref{eq:smKummerFE9}) is defined by $W\sim w_{*} \sim {\rm min}(1,  B^{-2})$ as suggested in the main text.  In this case the variable $\eta\sim 1/(1+A/\sqrt{v})$ is determined by $v \gg A^2$ so it can be set equal to unity. 

Hence, in the case of  $A^{-2} \ll $ min$(1, B^2)$ one can set $\eta(v)=1$ since the term $A\sqrt{w}$ is negligible compared to $w$ in the denominator of the integrand in Eq. (\ref{eq:smKummerFE8}). 
In the opposite asymptotic limit of min$(1, B^2) \ll A^{-2}$ one should set $\eta=1/2$ since the integral in Eq. (\ref{eq:smKummerFE8}) is determined by $w-A^2\ll A^2$. In both cases the assumption $\eta=const$ brings Eq.  (\ref{eq:smKummerFE}) to the hypergeometric form.

Consider the solution of Eq.  (\ref{eq:smKummerFE}) when the proper constant $\eta$ is chosen as described above. 
Then the  solutions can be expressed in terms of confluent hypergeometric functions \cite{AbramowitzBook,GradshteynRyzhik07}. Two linearly independent solutions with different asymptotic behaviors at zero and infinity are expressed by functions $M(a, b, v)$ and $U(a, b, v)$ \cite{Tricomi1947}. It is convenient to use the integral representation of these functions in the form \cite{AbramowitzBook,GradshteynRyzhik07}
\begin{eqnarray}
M(a, b, v) =  \frac{\Gamma(b)}{\Gamma(a)\Gamma(b-a)}\int_{0}^{1}e^{vt}t^{a-1}(1-t)^{b-a-1}dt, ~
U(a, b, v)=\frac{1}{\Gamma(a)}\int_{0}^{\infty}e^{-vt}t^{a-1}(1+t)^{b-a-1}dt, 
\label{eq:smKummerSol}
\end{eqnarray}
with the dimensionless parameters $a$ and $b$ defined as
\begin{eqnarray}
a=\frac{\eta B^2}{4}, ~ b=0. 
\label{eq:smKummerSol4}
\end{eqnarray}
The first solution approaches infinity in the limit $v \rightarrow \infty$, so only the second solution is acceptable. Correspondingly, the function $F$ can be represented as $cU(a, b, v)$ with the constant coefficient $c$ determined by the boundary condition, Eq. (\ref{eq:smKummerFB}). Applying this condition we get 
\begin{eqnarray}
F(v)=\frac{\pi\Omega_{R}U(\eta B^2/4, 0, v)}{U(\eta B^2/4, 0, A^2)-2\frac{dU(\eta B^2/4, 0, A^2)}{dA^2}}.
\label{eq:smKummerSol1}
\end{eqnarray}
Using the identity $dU(a, b, v)/dv=-aU(a+1, b+1, v)$ \cite{AbramowitzBook}  one can rewrite this solution as 
\begin{eqnarray}
F(v)=\frac{\pi\Omega_{R}U(\eta B^2/4, 0, v)}{U(\eta B^2/4, 0, A^2)+\frac{\eta B^2}{2}U(1+\eta B^2/4, 1, A^2)}.
\label{eq:smKummerSol2}
\end{eqnarray}
The loss tangent is determined by the solution $F(A^2)$ that can be expressed as 
\begin{eqnarray}
F(A^2)=\frac{\pi\Omega_{R}}{1-\frac{d\ln(U(a, b, A^2))}{dv}}=\frac{\pi\Omega_{R}}{1+\frac{\eta B^2}{2}\frac{U(\eta B^2/4, 0, A^2)}{U(1+\eta B^2/4, 1, A^2)}}.
\label{eq:smKummerSol3}
\end{eqnarray}

As argued above  in the case $A^2 \ll $ min$(1, B^{-2})$ one can set $\eta=1$. 
Then one can use the asymptotic behavior of the hypergeometric functions in this limit, given by \cite{AbramowitzBook}
\begin{eqnarray}
U(B^2/4, 0, A^2) \approx \frac{1}{\Gamma(1+B^2/4)}, ~ U(1+B^2/4, 1, A^2) \approx \frac{\ln(A^{-2})-2\gamma-\psi(1+B^2/4)}{\Gamma(1+B^2/4)}, 
\label{eq:smKummerAsympt}
\end{eqnarray}
where $\gamma\approx 0.5772$ is the Euler's constant and $\psi(x)=d\ln(\Gamma(x))/dx$ is the digamma function. The digamma function approaches $-\gamma$ for $B\rightarrow 0$ or it can be approximated by $2\ln(B/2)$ for $B \gg 1$ \cite{AbramowitzBook}. For the sake of simplicity we approximate it by the expression $\psi(1+B^2/4) \approx 2\ln(e^{-\gamma/2}+B/2)$, which is  valid in both limiting regimes. Consequently, the asymptotic behavior of Eq. (\ref{eq:smKummerAsympt}) in the case $A<1$ and $AB<1$ can be expressed as 
\begin{eqnarray}
F(A^2)\approx \frac{\pi\Omega_{R}}{1+B^2\ln\left(\frac{2e^{-\gamma/2}}{A(2e^{-\gamma}+B)}\right)}. 
\label{eq:smKummerSol2a}
\end{eqnarray}
As shown in Fig. \ref{fig:NonLinIntegr}, this solution  is in excellent agreement with the exact numerical solution of Eq. (\ref{eq:smKummerFE}) if the argument of the logarithm, that is supposed to be much greater then $1$, exceeds $2$.  This result is given in the main text in  Eq. 17. For the loss tangent calculations we used only its asymptotic behavior for $B > 1$ as described below.

\begin{figure}[!ht]
\begin{center}
\captionsetup[subfloat]{labelfont=bf}$
\begin{array}{cc}
\subfloat[]{\includegraphics[width=7cm]{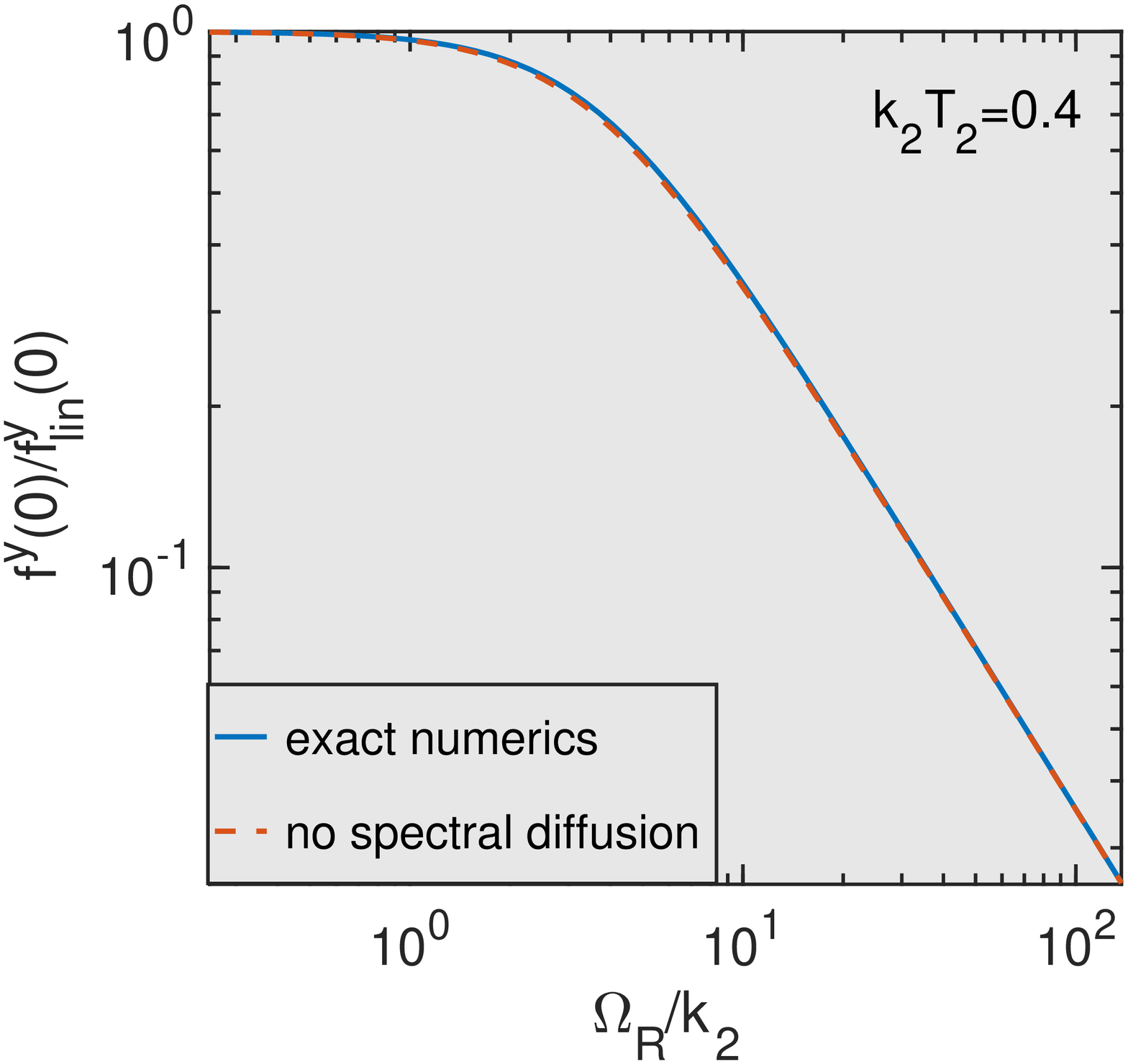}}&
\subfloat[]{\includegraphics[width=7cm]{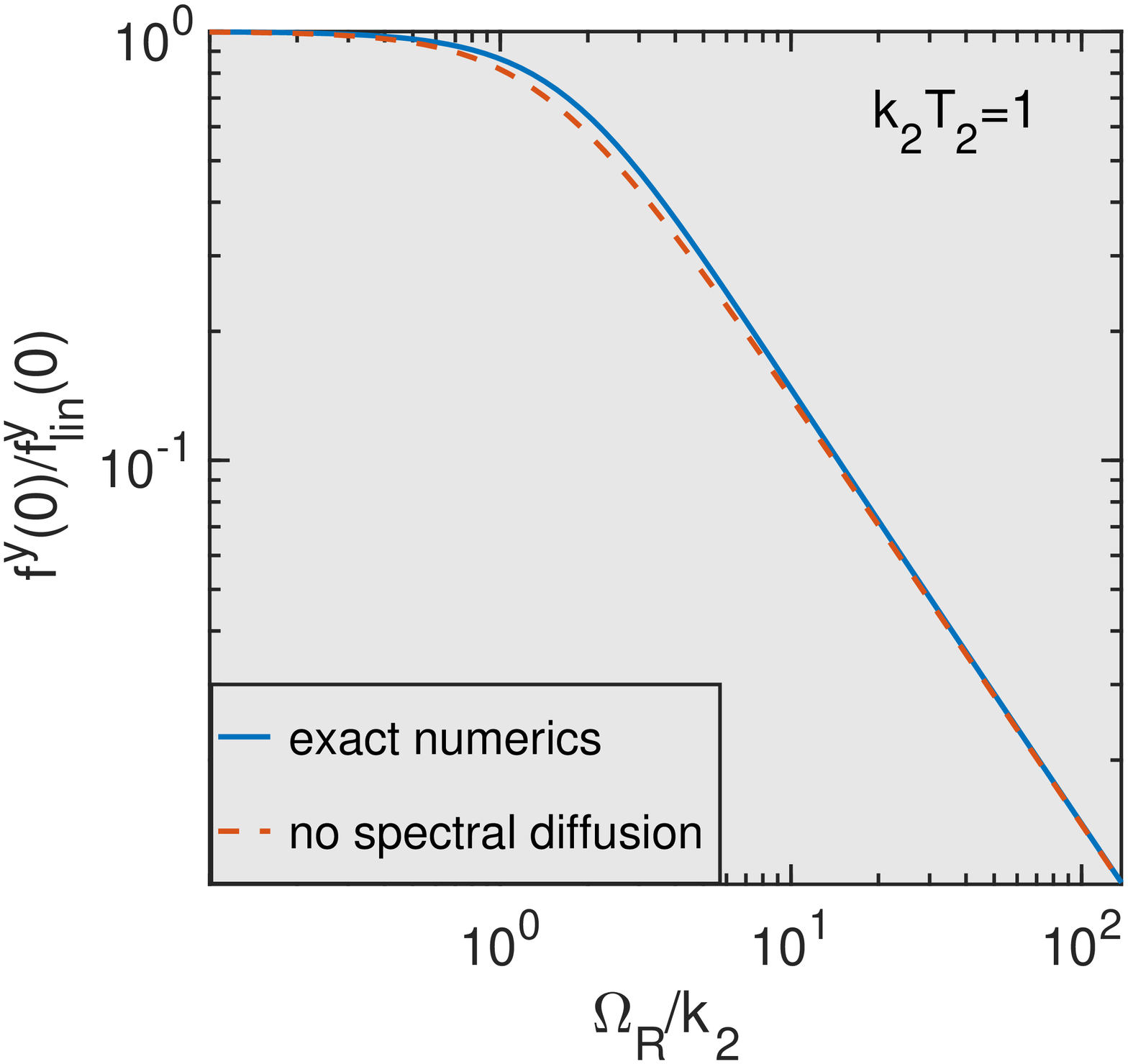}}\\
\subfloat[]{\includegraphics[width=7cm]{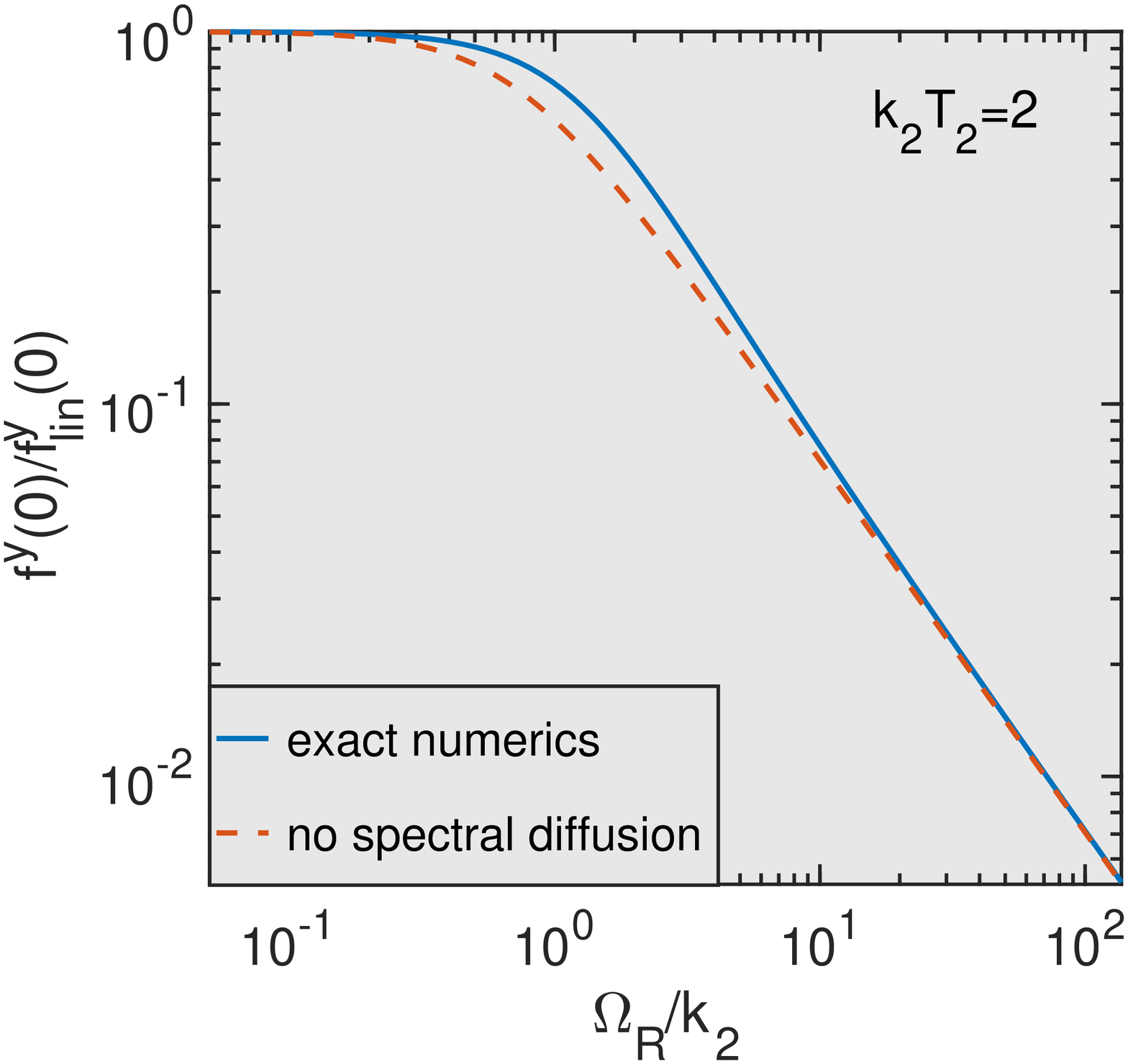}}&
\subfloat[]{\includegraphics[width=7cm]{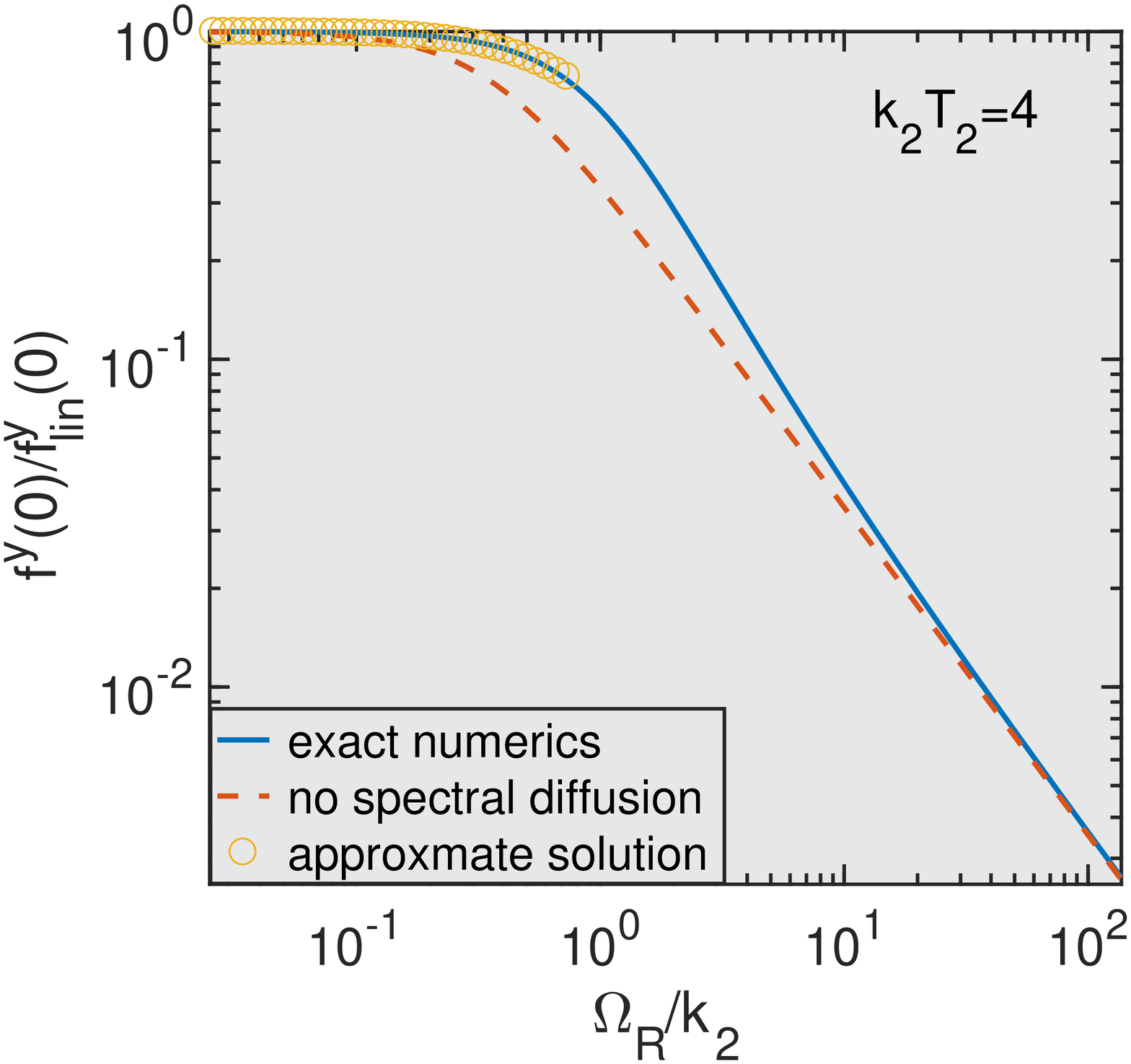}}\\
\subfloat[]{\includegraphics[width=7cm]{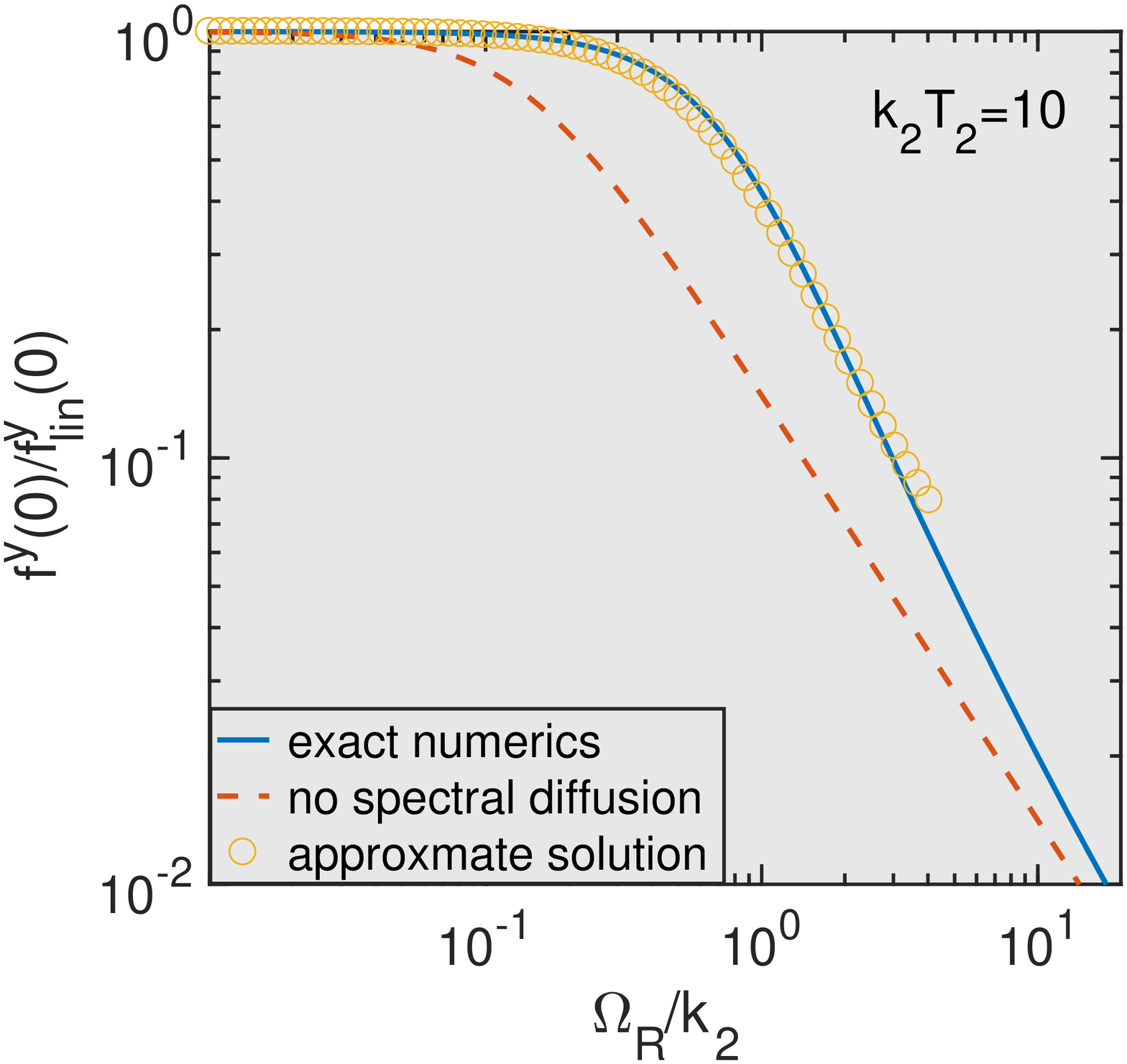}}&
\subfloat[]{\includegraphics[width=7cm]{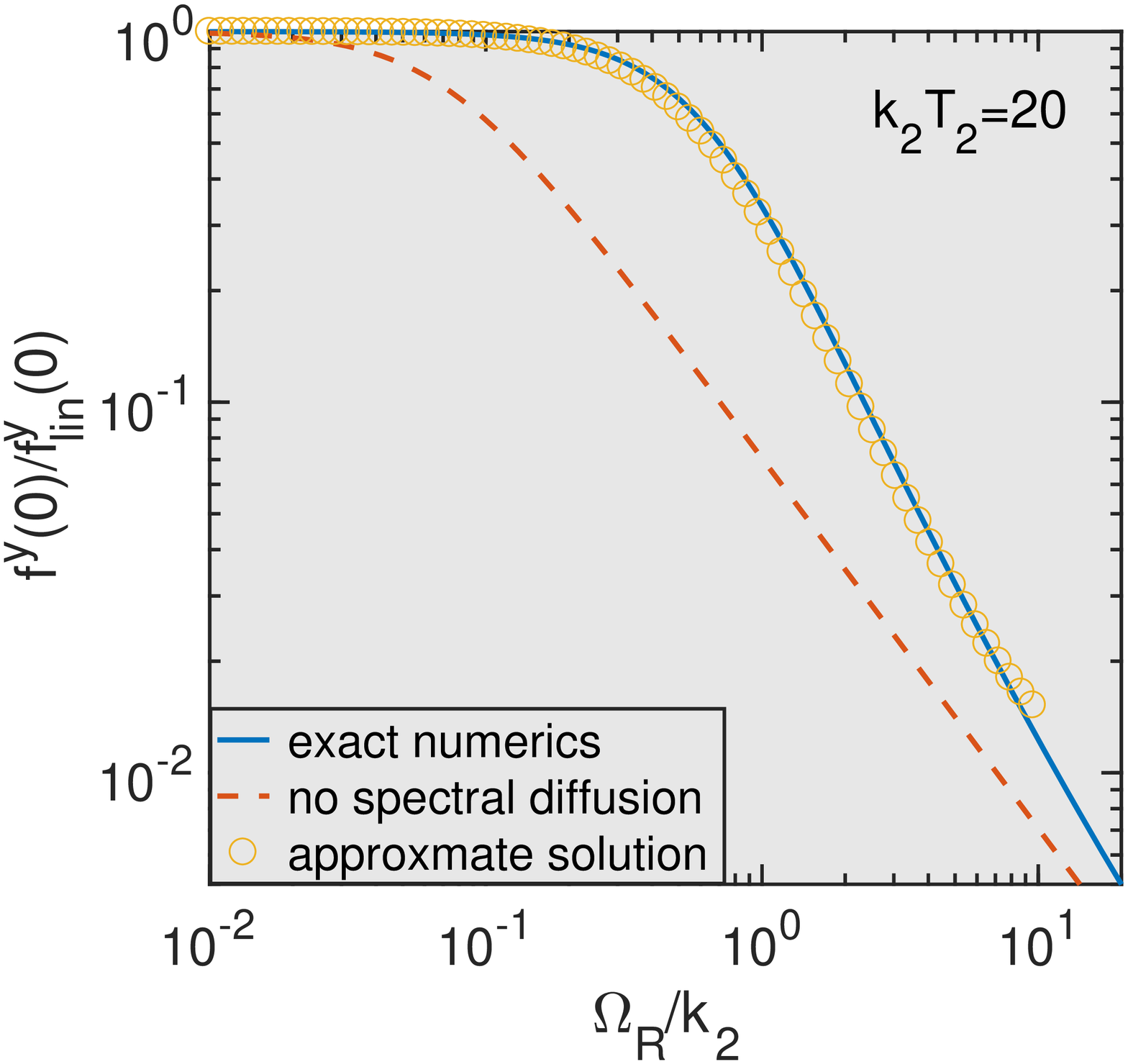}}
\end{array}$
\end{center}
\caption{ Comparison of numerical solution of Eq. (\ref{eq:smKummerFE}) and its analytical interpolations (solid line and hollow markers) for detuning averaged nonlinear absorption $f^{y}(0)/f^{y}_{lin}(0)$ ($f^{y}_{lin}(0)=\pi\Omega_{R}$ corresponds to the linear response theory limit) shown as a function of  Rabi frequency  for different relationships of spectral diffusion and relaxation rates $A^{-1}=k_{2}T_{2}$ ranging from $0.5$ (a) to $20$ (d). Dashed lines referred to as ``no spectral diffusion" show the solution of rate equations ignoring the spectral diffusion, Eq. (\ref{eq:smKummerSol3a}), and circles show the approximate solution, Eq. (\ref{eq:smKummerSol2a}), where it is applicable, i. e. the argument of logarithm exceeds $2$.}
\label{fig:NonLinIntegr}
\end{figure}

Consider the opposite limit $A^2 \gg {\rm min} (1, B^{-2})$. In that case one can approximately set $\eta=1/2$ (cf. Eq. (\ref{eq:smKummerFE})).
The asymptotic behavior of the confluent hypergeometric function $U$  given in Ref. \cite{AbramowitzBook} for this specific limit is complicated and contains sign variable series. However, the  logarithmic derivative of this function can be evaluated applying the saddle point approximation to the integral representation of Eq. (\ref{eq:smKummerSol}), since the integrand possesses a sharp maximum. Representing the integrand for the function $U$ in Eq. (\ref{eq:smKummerSol}) ($a=B^2/8$, $b=0$, $v=A^2$) as $e^{S(t)}$ with $S(t)=-vt+(a-1)\ln(t)-(a+1)\ln(1+t)$, one can find the position of this maximum solving the equation $dS/dt=0$. The solution can be written as $t_{max}=-1/2-A^{-2}+\sqrt{1/4+B^2/(8A^2)+1/A^4}$, and the logarithmic  derivative of interest can be estimated as $dS(t=t_{max})/dv=t_{max}$. In the case under consideration, one can ignore $A^{-2}$ and $A^{-4}$ terms compared to either 
$1/4$ or $(B/A)^{2}$ terms. Then the expression for the solution of interest $F(A^2)$ takes the standard form 
 \begin{eqnarray}
f^{y}(0)=F(A^2)\approx \frac{\pi\Omega_{R}}{\sqrt{1+\frac{B^2}{2A^2}}}=\frac{\pi\Omega_{R}}{\sqrt{1+\Omega_{R}^2T_{1}T_{2}}}.  
\label{eq:smKummerSol3a}
\end{eqnarray}
The same result can be obtained using the quasi-stationary solution of Eq. (\ref{eq:smKummerFE5}) with the derivative in its left hand side set to zero. 
This result  coincides with the stationary solution $\Omega_{R}\sigma^{y}$ of the system of rate equations for TLS density matrix \cite{Yu94} integrated over detuning. Integration of Eq. (\ref{eq:smKummerSol3a}) over TLS tunneling amplitudes and orientations with respect to the external field (Eq. 18 in the main text) leads to the familiar result for the loss tangent (Eq. 10 in the main text).

\section{Analytical and numerical evaluations of the loss tangent}

Here we describe in more detail the analytical calculation of the loss tangent in the case of the significant spectral diffusion where $T_{sd}/T_{2} \ll 1$ (see Eqs. 3 and 7 in the main text for the definitions of parameters). We begin with the consideration of intermediate Rabi frequencies $1/T_{sd}<\Omega_{R0}<T_{20}/T_{sd}^2$, where the non-linear microwave absorption by individual TLSs is determined by Eq. (\ref{eq:smKummerSol2a}). At smaller Rabi frequencies, $\Omega_{R}\ll 1/T_{sd}$, absorption takes place in the linear regime, while at larger Rabi frequency, $\Omega_{R0}>T_{20}/T_{sd}^2$, the detuning by the spectral diffusion $T_{20}/T_{sd}^2$ can be neglected compared to the Rabi frequency, and the loss tangent is defined by Eq. 4 in the main text (see also Eq. (\ref{eq:smKummerSol3a})). 

The loss tangent is determined by the integral of the product $\Omega_{R}f^{y}(0)$ (see Eq. 4 in the main text), while $f^{y}(0)=F(A^2)$ (see the definition of the function $F$ before Eq. (\ref{eq:smKummerFE})). In the case of interest this product can be written in the approximate analytical form, derived from Eq. (\ref{eq:smKummerSol2a}), 
\begin{eqnarray}
\Omega_{R}f^{y}(0)=\frac{\pi\Omega_{R}^2}{1+\frac{\Omega_{R}^2}{k_{2}^2} \ln\left(\frac{k_{2}^2T_{2}}{\Omega_{R}}\right)},
\label{eq:smIntegrExpr}
\end{eqnarray} 
where the numerical factors are skipped in the argument of the logarithm because their contribution is negligible compared to the large factor $k_{2}^2T_{2}/\Omega_{R}$ and the hypertangent factor in the numerator is set to unity here and below because the consideration is restricted  to low temperatures $T<\hbar\omega/k_{B}$ (cf Eq. 1 in the main text). The consideration is also limited to low temperatures $T<\hbar\omega/k_{B}$ so the population difference expressed by the hypertangent factor is set to unity (cf Eqs. 2, 5 and 6 in the main text).  
The loss tangent is determined by the weighted integral (Eq. 4 in the main text) of Eq. (\ref{eq:smIntegrExpr}) over TLS parameters $x$ and $y$ where $x=\Delta_{0}/E$ is the ratio of TLS  tunneling amplitude and energy, while $y=\cos(\theta)$ stands for the cosine of the angle between the TLS dipole moment $\mathbf{p}$ and the external field $\mathbf{F_{AC}}$. Using the definitions of TLS parameters (Eqs. 3 and 7 in the main text), one can express $x$ and $y$ dependencies of $A$ and $B$ as $A=1/(k_{2}T_{2})=(x^2/(1-x^2)^{1/4})T_{sd}/T_{20}$ and $B=\Omega_{R}/k_{2}=(xy/(1-x^2)^{1/4})\Omega_{R0}T_{sd}$. Then the approximate analytical expression for the loss tangent can be written as  (the contribution of a very narrow domain $1-x \ll 1$ is negligible and can be ignored)
\begin{eqnarray}
\tan(\delta)=\frac{4\pi^2 P_{0}\hbar^2 \Omega_{R0}^2}{2\epsilon'F_{AC}^2} \int_{-1}^{1}dy\int_{0}^{1}dx \frac{x y^2}{\sqrt{1-x^2}}\frac{1}{1+\frac{x^2y^2(\Omega_{R0}T_{sd})^2}{\sqrt{1-x^2}} \ln\left(\frac{T_{20}(1-x^2)^{1/2}}{x^3y\Omega_{R0}T_{sd}^2}\right)}.
\label{eq:smlossdef}
\end{eqnarray}

In the  case of a small external field, $\Omega_{R0}T_{sd} \ll 1$, one has $\Omega_{R}f^{y}(0)\approx \pi\Omega_{R}^2$, and the loss tangent takes the standard form of the linear response theory (cf. Eq. 6 in the main text)
\begin{eqnarray}
\tan(\delta_{0})=\frac{4\pi^2 P_{0}\hbar^2\Omega_{R0}^2}{2\epsilon'F_{AC}^2} \int_{-1}^{1}y^2dy\int_{0}^{1} dx\frac{x}{\sqrt{1-x^2}}=\frac{4\pi^2 P_{0}p^2}{3\epsilon'}.
\label{eq:smlossdefLR}
\end{eqnarray}

In the most interesting regime of $1 \ll \Omega_{R0}T_{sd} \ll T_{20}/T_{sd}$, the integral in Eq. (\ref{eq:smlossdef}) with respect to $x$ is essentially logarithmic. This permits us to evaluate it within logarithmic accuracy. Assuming the logarithmic accuracy one can replace the factor $(1-x^2)^{1/2}$ and the variable $y$  in the integrand by some ``acting" values $z_{*} \sim 1$  and $y_{*} \sim 1$.
Then Eq. (\ref{eq:smlossdef}) can be rewritten as 
\begin{eqnarray}
\tan(\delta)=\frac{4\pi^2 P_{0}\hbar^2 \Omega_{R0}^2}{\epsilon'F_{AC}^2} \int_{0}^{1}dx \frac{y_{*}^2 x}{z_{*}}\frac{1}{1+\frac{x^2y_{*}^2(\Omega_{R0}T_{sd})^2}{z_{*}} \ln\left(\frac{T_{20}z_{*}}{\sqrt{2}x^3y_{*}\Omega_{R0}T_{sd}^2}\right)}.
\label{eq:smlossEval1}
\end{eqnarray}
To evaluate this integral within logarithmic accuracy, one can consider only its entirely logarithmic part where the second term in the denominator exceeds $1$. This can be done by setting the lower integration limit to $x_{min}=\eta_{*}/(\Omega_{R0}T_{sd})$, ignoring $1$ in the denominator, and using $\eta_{*} \sim 1$ as the adjustment parameter to account for the omitted domain. Then the integral takes the form
\begin{eqnarray}
\tan(\delta)=\frac{4\pi^2 P_{0}\hbar^2}{\epsilon'F_{AC}^2 T_{sd}^2} \int_{\frac{\eta_{*}}{\Omega_{R0}T_{sd}}}^{1} \frac{dx}{x\ln\left(\frac{T_{20}z_{*}}{\sqrt{2}x^3y_{*}\Omega_{R0}T_{sd}^2}\right)}.
\label{eq:smlossEval2}
\end{eqnarray}
This integral can be evaluated analytically using the substitution $z=\ln(x)$ and the result of the integration reads 
\begin{eqnarray}
\tan(\delta)=\frac{4\pi^2 P_{0}\hbar^2}{3\epsilon'F_{AC}^2 T_{sd}^2} \ln\left[1+\frac{3\ln\left(\Omega_{R0}T_{sd}/\eta_{*}\right)}{\ln\left(\frac{T_{20}z_{*}}{\sqrt{2}y_{*}\Omega_{R0}T_{sd}^2}\right)}\right]=\frac{\tan(\delta_{0})}{\Omega_{R}^2T_{sd}^2}\ln\left[1+\frac{3\ln\left(c_{1}\Omega_{R0}T_{sd}\right)}{\ln\left(\frac{c_{2}T_{20}}{\Omega_{R0}T_{sd}^2}\right)}\right], ~ c_{1}=\frac{1}{\eta_{*}}, ~ c_{2}=\frac{z_{*}}{\sqrt{2}y_{*}}. 
\label{eq:smlossEval3}
\end{eqnarray}
In the asymptotic regime  $1/T_{sd}\ll \Omega_{R0}\ll T_{20}/T_{sd}^2$ the numerical parameters $c_{1}$ and $c_{2} \sim 1$ are negligible in the arguments of logarithms compared to the large factors $\Omega_{R0}T_{sd}$ and $T_{20}/(\Omega_{R0}T_{sd}^2)$ so they can be skipped. However, they can be important in the crossover regime as described below. 

In the case of a very large external field, $T_{20}/T_{sd} \ll \Omega_{R0}T_{sd}$ one can use Eq. (\ref{eq:smKummerSol3a}). Then, ignoring unity in the denominator compared to the second term $\Omega_{R}T_{2} \gg 1$, we get 
\begin{eqnarray}
\tan(\delta)=\frac{4\pi^2 P_{0}\hbar^2 \Omega_{R0}^2}{\epsilon'F_{AC}^2} \int_{0}^{1}dy\int_{0}^{1} \frac{\sqrt{2}x^2 ydx}{\sqrt{1-x^2}}
\frac{1}{\Omega_{R0}T_{20}}=\frac{4\pi^2 P_{0}\hbar^2 \Omega_{R0}^2}{\epsilon'F_{AC}^2}\frac{\pi}{4\sqrt{2}\Omega_{R0}T_{20}}=\tan(\delta_{0})\frac{3\pi}{4\sqrt{2}\Omega_{R0}T_{20}}.
\label{eq:smlossEval4}
\end{eqnarray}
The latter result describes non-linear regime in the absence of the spectral diffusion (Eq. 10 in the main text). 

It is useful practically to have a single  analytical interpolation matching both behaviors of Eq. (\ref{eq:smlossEval3}) and Eq. (\ref{eq:smlossEval4}) and the crossover between them. This is attained combining two results as 
\begin{eqnarray}
\tan(\delta_{sd})=\frac{\tan(\delta_{0})}{1+
\frac{\Omega_{R0}^2T_{sd}^2}{\frac{3\pi \sqrt{2}\Omega_{R0}T_{sd}^2}{8T_{20}}+\ln(1+3l_{*})}}, ~
l_{*}=\frac{\ln\left(d_{1}+c_{1}\Omega_{R0}T_{sd}\right)}{
\ln\left(d_{2}+c_{2}\frac{T_{20}}{\Omega_{R0}T_{sd}^2
}\right)},
\label{eq:smlossinterp}
\end{eqnarray}
where the adjustable parameters $d_{1}$ and $d_{2}$  are introduced to provide a ``reasonable" behavior of the function in the case of small arguments under logarithms to keep them positive.  Consequently, parameters $d_{1}$ and $d_{2}$ should not be less than unity.  Two other adjustable parameters $c_{1}$ and $c_{2}$ are determined by the uncertainty of calculations within  logarithmic accuracy expressed through unknown parameters  $y_{*}$, $z_{*}$ and $\eta_{*}$ in Eq. (\ref{eq:smlossEval3}). We expect that all these parameters are of order of $1$. The optimum Monte-Carlo fit of the numerical solution by interpolating Eq. (\ref{eq:smlossinterp}) (cf. Ref. \cite{ab13echo}) results in the following definitions of parameters: $c_{1} \approx 0.5$, $c_{2} \approx 0.55$, $d_{1} \approx 3$, $d_{2} \approx 1.2$. This interpolation formula is given in Eqs. 19, 20 in the main text. In the case $T_{20}/T_{sd}>5$ this equation fits the exact solution with the accuracy of around $1.5\%$ while for $T_{20}/T_{sd}\leq 1$ it is not quite applicable and the earlier result (Eq. 6 in the main text) should be used, see Figs. \ref{fig:NonLinIntegr1} a-f. 

\begin{figure}[!ht]
\begin{center}
\captionsetup[subfloat]{labelfont=bf}$
\begin{array}{cc}
\subfloat[]{\includegraphics[width=7cm]{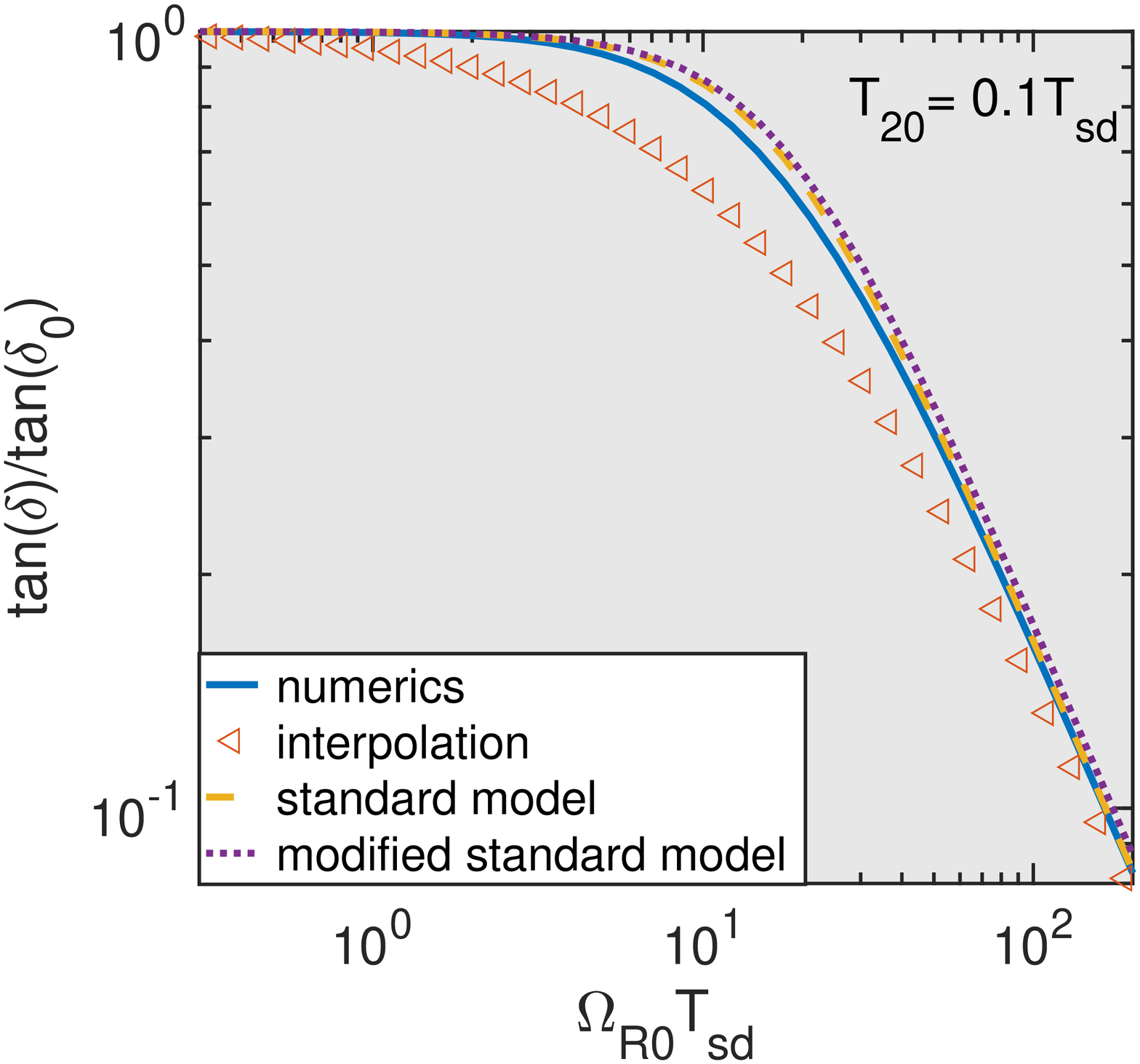}}&
\subfloat[]{\includegraphics[width=7cm]{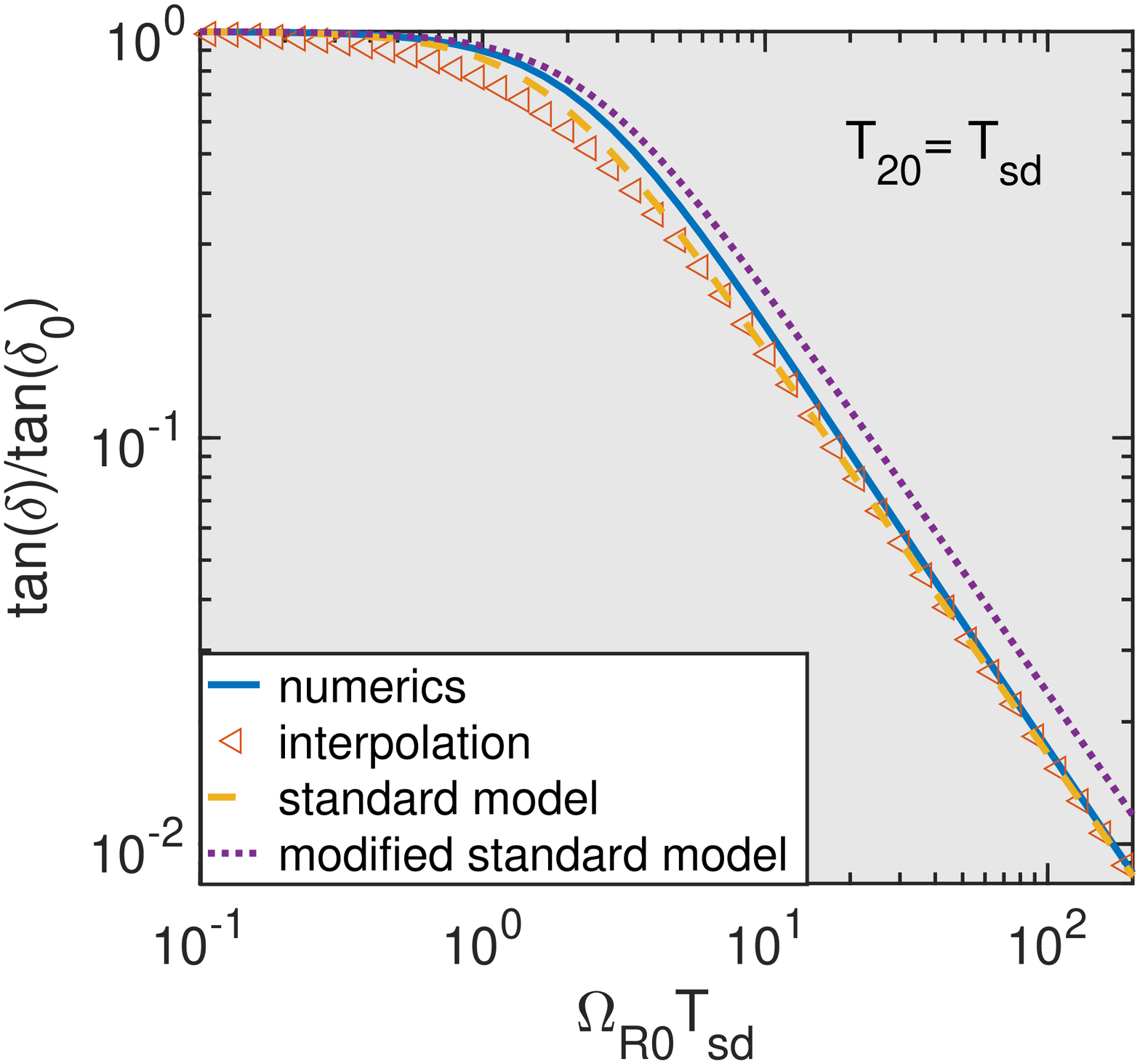}}\\
\subfloat[]{\includegraphics[width=7cm]{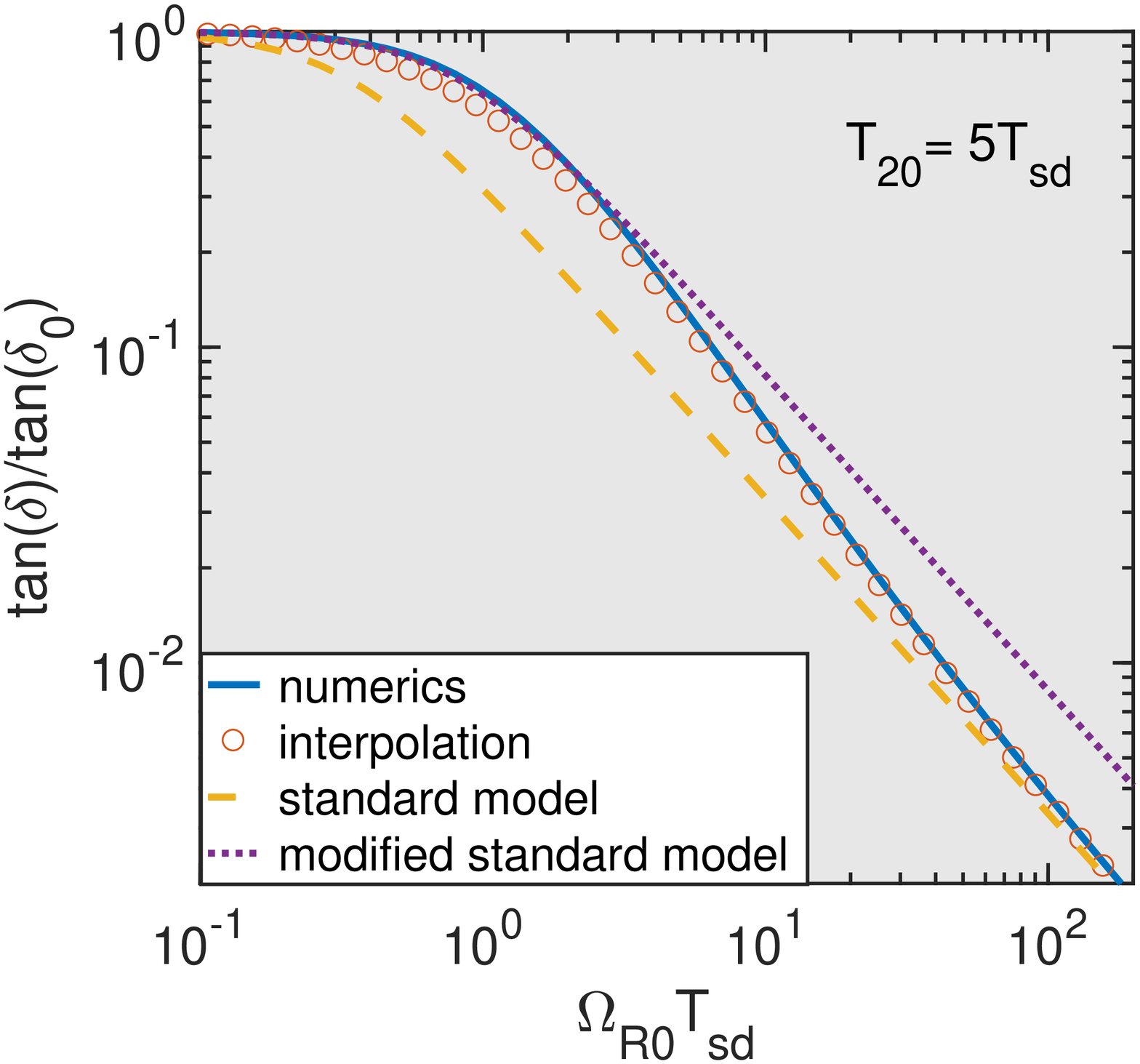}}&
\subfloat[]{\includegraphics[width=7cm]{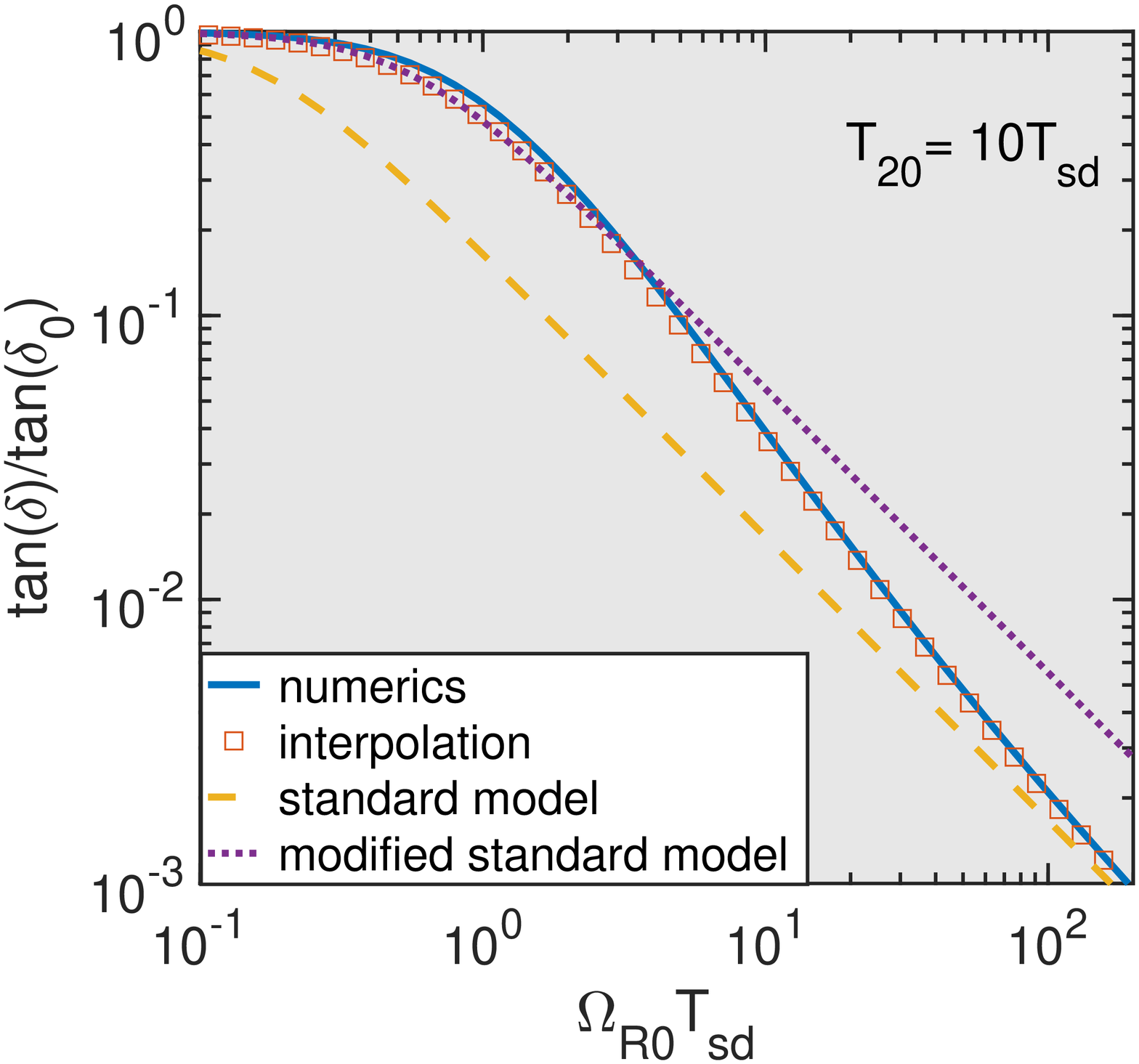}}\\
\subfloat[]{\includegraphics[width=7cm]{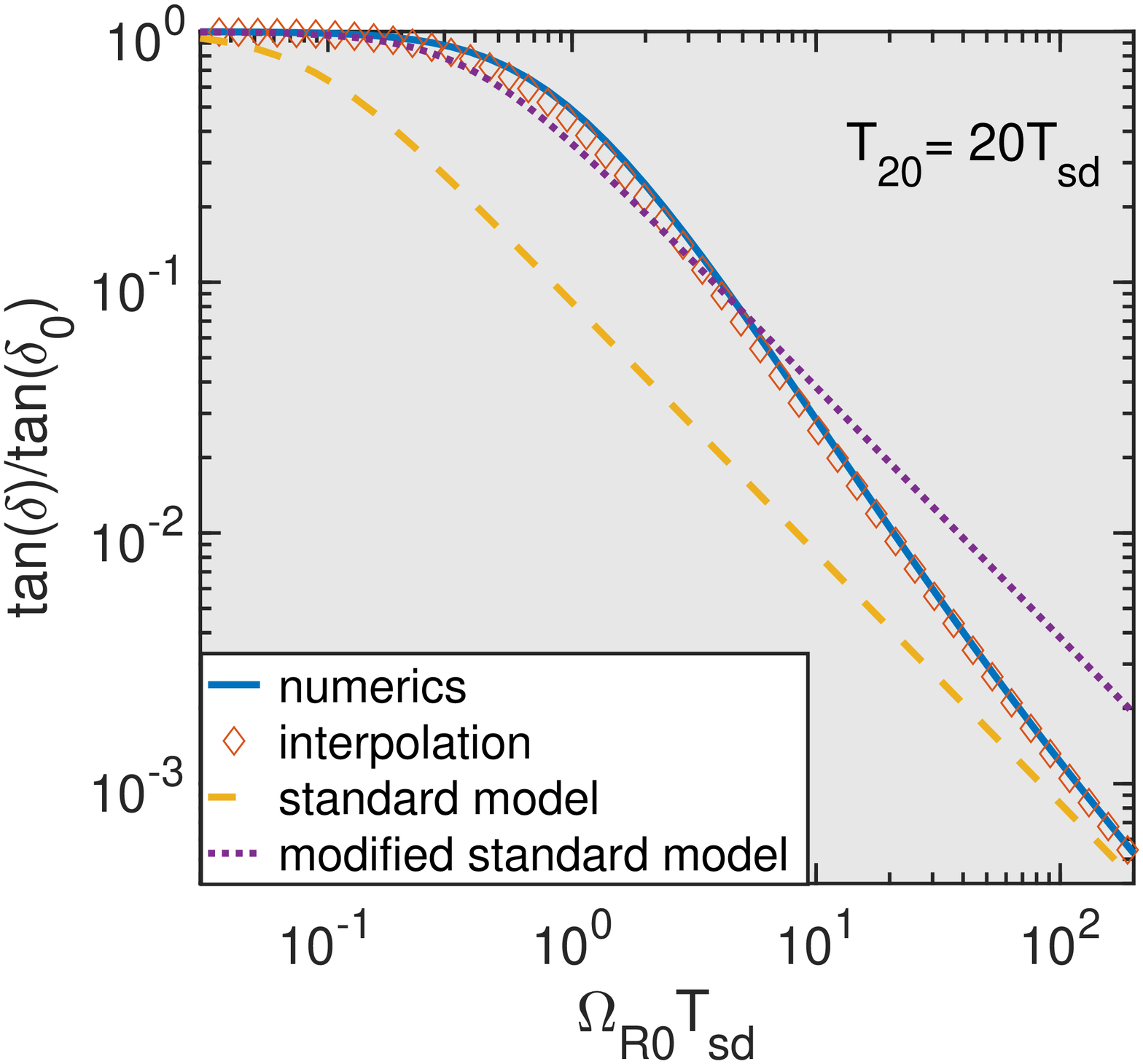}}&
\subfloat[]{\includegraphics[width=7cm]{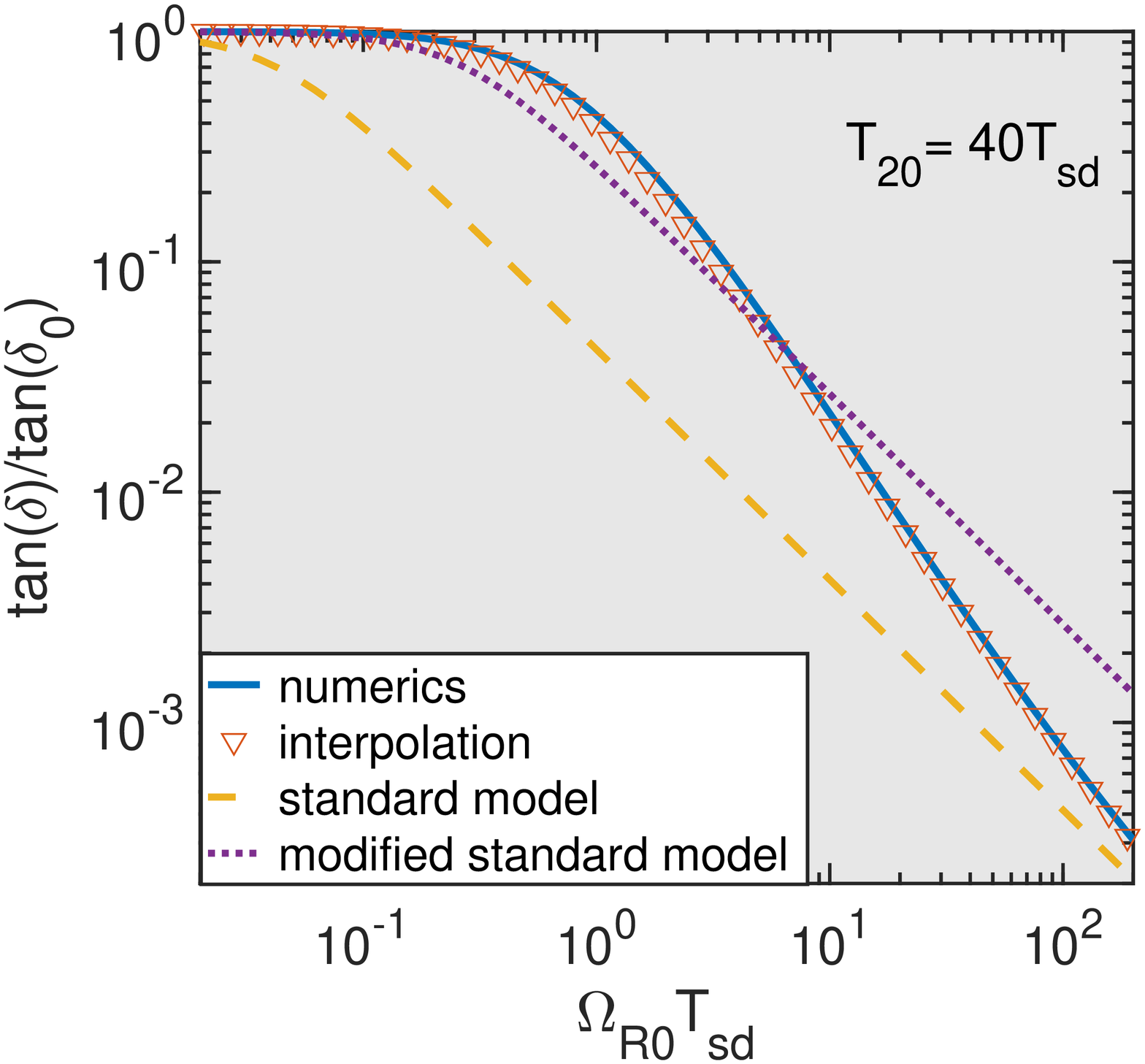}}
\end{array}$
\end{center}
\caption{ Numerical and analytical solutions for non-linear absorption dependence on the maximum Rabi frequency for different relationships of phase decoherence and relaxation times $T_{20}/T_{sd}$ ranging from $0.1$ (a) to $40$ (e). Dashed and dotted lines show the solution of rate equations ignoring the spectral diffusion (standard model) and including its rate directly to the decoherence rate as described in the main text (modified standard model), respectively.}
\label{fig:NonLinIntegr1}
\end{figure}

\end{widetext}

\end{document}